\documentclass[twocolumn,twocolappendix]{aastex701}

\usepackage{graphicx}
\usepackage{amsmath}
\usepackage{siunitx}
\usepackage{natbib}
\usepackage{hyperref}
\usepackage{mathptmx}

\newcommand\mean{mean}
\newcommand\median{median}
\newcommand\std{sd}
\newcommand\dphot{d$_{\mathrm{phot}}$}
\newcommand\ms{$-$}
\newcommand\ps{$+$}

\shorttitle{Late-type brown dwarfs in the Euclid Q1 Release}
\shortauthors{Kiwy et al.}

\begin{document}

    \title{A search for late-type brown dwarfs in the Euclid Quick Data Release 1}

    \author[0000-0001-8662-1622]{Frank Kiwy}
    \affiliation{Backyard Worlds: Planet 9}
    \email{frank.kiwy@outlook.com}

    \author[0000-0003-4269-260X]{J. Davy Kirkpatrick}
    \affiliation{IPAC, Mail Code 100--22, Caltech, 1200 E. California Blvd., Pasadena, CA 91125, USA}
    \email{davy@ipac.caltech.edu}

    \author[0000-0002-6294-5937]{Adam C. Schneider}
    \affiliation{United States Naval Observatory, Flagstaff Station, 10391 West Naval Observatory Rd., Flagstaff, AZ 86005, USA}
    \email{adam.c.schneider4.civ@us.navy.mil}

    \author[0000-0002-1125-7384]{Aaron M. Meisner}
    \affiliation{NSF National Optical-Infrared Astronomy Research Laboratory, 950 N. Cherry Ave., Tucson, AZ 85719, USA}
    \affiliation{Carl and Lily Pforzheimer Foundation Fellow, Radcliffe Institute for Advanced Study at Harvard University, 10 Garden Street, Cambridge, MA 02138, USA}
    \affiliation{Center for Astrophysics $|$ Harvard \& Smithsonian, 60 Garden St., Cambridge, MA 02138, USA}
    \email{aaron.meisner@noirlab.edu}

    \author[0000-0001-6251-0573]{Jacqueline K. Faherty}
    \affiliation{Department of Astrophysics, American Museum of Natural History, Central Park West at 79th Street, NY 10024, USA}
    \email{jfaherty@amnh.org}

    \author[0000-0002-2387-5489]{Marc J. Kuchner}
    \affiliation{Exoplanets and Stellar Astrophysics Laboratory, NASA Goddard Space Flight Center, 8800 Greenbelt Road, Greenbelt, MD 20771, USA}
    \email{marc.j.kuchner@nasa.gov}

    \author[0000-0001-8170-7072]{Daniella Bardalez Gagliuffi}
    \affiliation{Department of Physics \& Astronomy, Amherst College, 25 East Drive, Amherst, MA 01003, USA}
    \email{daniellabardalez@gmail.com}

    \author[0000-0003-2478-0120]{Sarah L. Casewell}
    \affiliation{School of Physics and Astronomy, University of Leicester, University Road, Leicester, LE1 7RH, UK}
    \email{slc25@leicester.ac.uk}

    \author[0000-0003-2235-761X]{Thomas P. Bickle}
    \affiliation{Backyard Worlds: Planet 9}
    \affiliation{School of Physical Sciences, The Open University, Milton Keynes, MK7 6AA, UK}
    \email{tombicklecrypto@gmail.com}

    \collaboration{all}{The Backyard Worlds: Planet 9 Collaboration}

    \begin{abstract}
        We present the identification and characterization of 15 mid-to-late T dwarf candidates in the Euclid Quick Release 1 (Q1) dataset, based on a combined photometric and spectroscopic analysis.
        Candidates were initially selected via color-based cuts in the Euclid $Y_E - J_E$ and $J_E - H_E$ color–color space, targeting the region occupied by ultracool dwarfs in synthetic photometry from the \citet{Sanghi+2024} sample.
        From an initial pool of 38,845 sources, we extracted low-resolution near-infrared spectra from the Euclid NISP instrument and applied a two-stage validation procedure that included spectral template fitting followed by visual inspection.
        Eight of the 15 validated candidates are newly identified objects with no prior literature association.
        We examined their morphological and photometric properties and compared them with established spectral standards.
        Photometric distances were derived using band-averaged distance modulus estimates.
        We discuss the limitations and promise of the Euclid survey for ultracool dwarf studies, and demonstrate the potential for discovering substellar populations beyond the reach of current wide-field surveys.
    \end{abstract}


    \section{Introduction}\label{sec:introduction}

    Brown dwarfs occupy the range between the highest mass giant planets and the lowest mass stars, typically between $\sim$13 and 80 Jupiter masses, lacking the core temperatures necessary to sustain hydrogen fusion~\citep{Hubbard+1997, Burrows+2001, Baraffe+2003, Moraux+2003, Spiegel+2011, Dieterich+2018, Grieves+2021}.
    As a consequence, these substellar objects cool and fade over time, making their detection increasingly challenging at later evolutionary stages and lower masses~\citep{Chabrier+2005, Kirkpatrick+2012}.
    The identification and characterization of brown dwarfs, particularly those of late spectral types (e.g., T and Y dwarfs), is essential for understanding the low-mass end of the initial mass function, substellar cooling rates, and the overall census of the solar neighborhood~\citep{Luhman+2000, Huston+2021, Kirkpatrick+2021, Kirkpatrick+2024, Best+2024}.

    Large-scale photometric and spectroscopic surveys have greatly expanded the known population of brown dwarfs over the last few decades (e.g. \citealt{Cushing+2011, Kirkpatrick+1999, Kirkpatrick+2021, Kirkpatrick+2024}).
    Ground-based efforts such as the Sloan Digital Sky Survey (SDSS; \citealt{York+2000}), the Two Micron All Sky Survey (2MASS; \citealt{Cutri+2003, Skrutskie+2006}), the UKIRT Infrared Deep Sky Survey (UKIDSS; \citealt{Lawrence+2007}), the VISTA Hemisphere Survey (VHS; \citealt{McMahon+2013}), and the Panoramic Survey Telescope and Rapid Response System Survey (Pan-STARRS1; \citealt{Chambers+2016}) have been instrumental in identifying large samples of L and early-T dwarfs.
    Space-based missions like the Wide-field Infrared Survey Explorer (WISE; \citealt{Wright+2010}) and Spitzer Space Telescope (Spitzer; \citealt{Werner+2004}) extended this reach into the mid-infrared, enabling the discovery of even cooler objects, including late-T and Y dwarfs.
    More recently, Gaia~\citep{Gaia+2016} has provided high-precision astrometry for brighter ultracool dwarfs, while the James Webb Space Telescope (JWST; \citealt{Gardner+2006}) has begun to probe the faintest and coldest members of the substellar population with unprecedented sensitivity and resolution (e.g. \citealt{Beiler+2024, Hainline+2024, Faherty+2024, Faherty+2025, Chen+2025}).
    These recent studies demonstrate that the coldest objects continue to reveal new and unexpected aspects of substellar atmospheres and chemistry.
    Taken together, these surveys now provide a robust framework for classifying and characterizing brown dwarfs across a wide range of temperatures and masses.

    The Euclid space telescope~\citep{Laureijs+2010}, although primarily designed for cosmology, offers a unique opportunity to identify faint brown dwarfs through its wide-field near-infrared photometry and slitless spectroscopy.
    Its Near-Infrared Spectrometer and Photometer (NISP; \citealt{EuclidNISP+2025}) provides simultaneous $Y_E$, $J_E$, and $H_E$ photometry along with low-resolution spectra (R $\simeq$ 500) over a contiguous near-infrared range (1.2–\SI{1.9}{\micro\metre}), making it particularly suitable for detecting the broad absorption features characteristic of T-type atmospheres.

    Our work complements several recent studies utilizing Euclid data.
    Notably, \citet{Dominguez-Tagle+2025} and \citet{Zerjal+2025} conducted spectroscopic and photometric searches for ultracool dwarfs (UCDs), respectively, while \citet{Mohandasan+2025} spectroscopically identified UCDs in the Deep Field North, primarily of earlier spectral types.
    More recently, \citet{ZJZhang+2025} used machine learning tools to photometrically identify 25 UCDs in the M and L regimes.
    In contrast, our study combines photometric selection with spectroscopic validation, focusing specifically on late-type brown dwarfs.

    In this work, we present a search for mid-to-late T dwarfs in the Euclid Q1 data release~\citep{EuclidQ1+2025}.
    We begin with a photometric selection in the $Y_E - J_E$ vs. $J_E - H_E$ color–color space to isolate potential candidates.
    These sources are then subjected to a two-stage validation process that combines automated spectral template fitting and manual inspection.
    A final sample of 15 high-confidence brown dwarf candidates is identified and analyzed in detail.

    This paper is outlined as follows.
    In Section~\ref{sec:photometric-selection}, we describe the color-based selection strategy.
    Section~\ref{sec:spectroscopic-validation} details the spectroscopic validation of the candidates.
    The properties of the final sample are examined in Section~\ref{sec:candidate-analysis}, including comparisons with spectral standards.
    In Section~\ref{sec:refinement-of-photometric-selection}, we propose refined selection criteria informed by our results.
    Section~\ref{sec:discussion} discusses our findings in the context of previous studies, classification uncertainties, and the implications for future Euclid searches.
    Finally, we summarize our findings in Section~\ref{sec:conclusions}.

    \section{Photometric Selection}\label{sec:photometric-selection}

    To isolate candidate late-type brown dwarfs in the Euclid Q1 dataset, we focused on the three near-infrared (NIR) filters—$Y_E$, $J_E$, and $H_E$—which are particularly well suited for detecting mid-to-late T dwarfs.

    These cool objects are significantly brighter in the infrared than at visual wavelengths, making the Euclid NISP bands ideal for their detection and characterization.
    In contrast, the Visual Imager (VIS; \citealt{EuclidVIS+2025}) is not well suited for this purpose: late-T dwarfs are intrinsically very faint at optical wavelengths, and are often undetected or detected at very low signal-to-noise (S/N) in the $VIS$-band, especially at the faint end of the Q1 sample.
    While some brighter late-T dwarfs may be detected by VIS, relying on it would lead to incompleteness and potential selection biases.

\begin{figure*}
        \begin{center}
            \includegraphics[width=15cm]{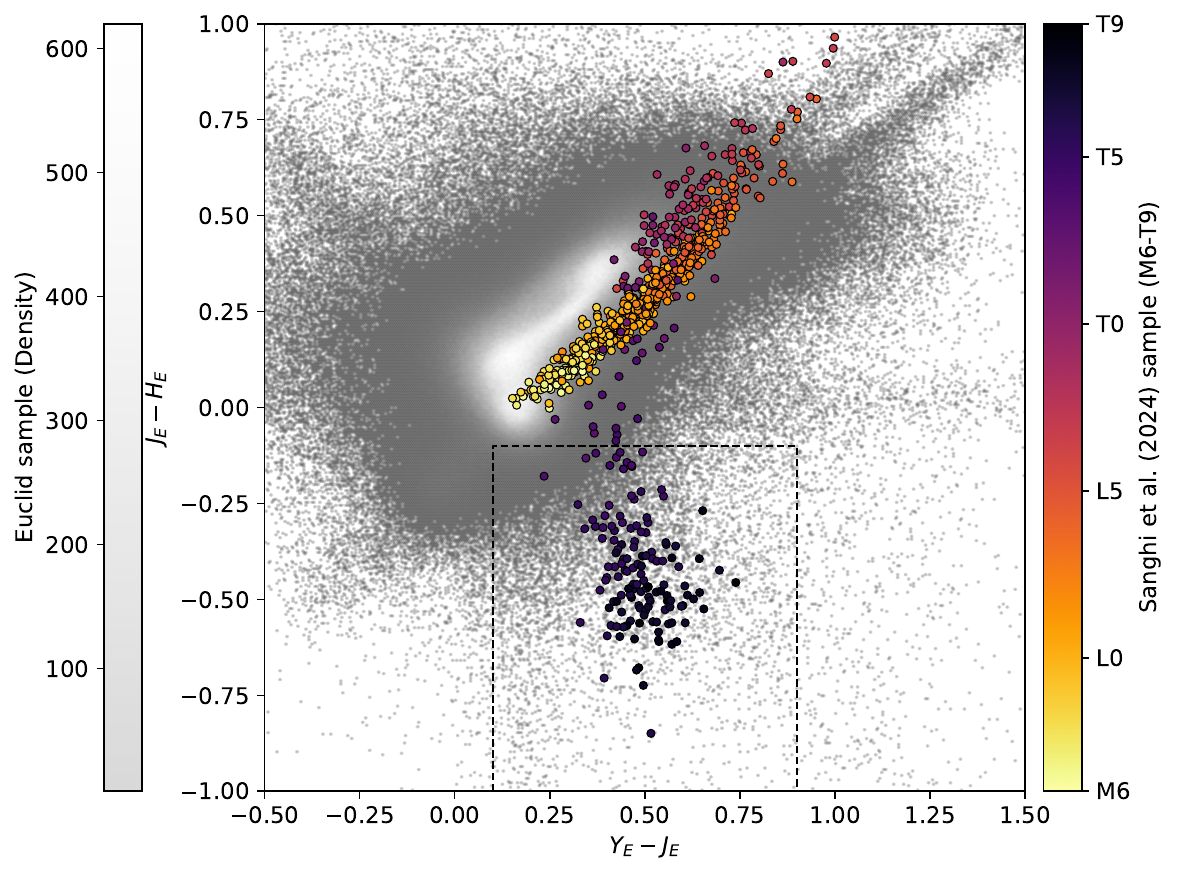}
        \end{center}
        \caption{
            Color–color diagram showing the Euclid $Y_E - J_E$ vs. $J_E - H_E$ color space. The grayscale background represents the density of sources from the Euclid Q1 dataset, limited to high-quality detections and magnitudes brighter than 23 mag in all three bands. Overplotted are synthetic Euclid colors of M6–T9 ultracool dwarfs from the \citet{Sanghi+2024} sample, colored by spectral type. Mid-to-late T dwarfs clearly deviate from earlier-type UCDs, forming a distinct population toward redder $Y_E - J_E$ and bluer $J_E - H_E$ colors. The black dashed lines indicate the adopted color selection boundaries ($0.1 < Y_E - J_E < 0.9$ and $J_E - H_E < -0.1$) used to identify candidate mid-to-late T dwarfs in the Euclid Q1 dataset.
        }
        \label{fig:sanghi_sample}
    \end{figure*}

    To define our selection region, we used synthetic photometry from the \citet{Sanghi+2024} sample of ultracool dwarfs, spanning spectral types M6 to T9, and mapped their locations in the Euclid color–color space defined by $Y_E - J_E$ and $J_E - H_E$.
    The background distribution in Figure~\ref{fig:sanghi_sample} shows the density of sources from the Euclid Q1 merged catalogue~\citep{EuclidMER+2025} with high-quality detections (\texttt{det\_quality\_flag = 0}) and positive fluxes in the $Y_E$, $J_E$, and $H_E$ bands.
    Magnitudes were computed from the template-fit fluxes using the AB zero-point, and we imposed an upper magnitude limit of 23 mag in all bands to ensure good photometric reliability and minimize contamination from faint, low-S/N sources.
    None of the 15 candidates in our final sample approach this limit, with the faintest having $Y_E$ = 22.5 mag.

    The \citet{Sanghi+2024} objects are shown in Figure~\ref{fig:sanghi_sample} with a spectral-type color gradient.
    Mid-to-late T dwarfs clearly separate from earlier-type ultracool dwarfs (late M, L, and early T) and from the bulk of the main stellar and galaxy loci.
    Based on this distribution, we adopted color constraints of $0.1 < Y_E - J_E < 0.9$ and $J_E - H_E < -0.1$, indicated by black dashed lines.

    The application of these color and magnitude constraints to the Euclid Q1 dataset yielded an initial sample of 38,845 sources that were retained for further analysis.

    \section{Spectroscopic Validation}\label{sec:spectroscopic-validation}

    While the color-based photometric selection yields 38,845 potential sources, the vast majority are contaminants.
    To refine this sample, we employed a two-stage validation procedure that combined spectral template fitting with visual inspection.

    In the first stage, we retrieved available spectra for each source from the ESA Euclid archive \citep{Euclid+2025} and compared them to standards defined by \citet{Burgasser+2006}, \citet{Kirkpatrick+2010} and \citet{Cushing+2011}, as compiled by~\citet{Burgasser+2017}, covering spectral types from M0 to T9.
    Each Euclid spectrum was trimmed to the 1.22–\SI{1.88}{\micro\metre} range to avoid increased noise toward the spectral edges.
    We chose to exclude only pixels with zero flux values and did not apply additional filtering based on the \texttt{MASK} (pixel mask), \texttt{QUALITY} (pixel quality flag), or \texttt{NDITH} (number of spectra used in combination) columns included in the spectrum FITS files.
    While more stringent filtering criteria are commonly employed, we found that such filtering may unnecessarily exclude data points that, despite minor calibration warnings, still provide useful information for spectral classification.
    A case in point is the spectrum of E0328\ms2749, shown in Figure~\ref{fig:comparison-with-burgasser-templates}, which yields a robust fit despite containing many flagged pixels (marked in orange).
    We therefore opted for a more inclusive approach to preserve potentially informative spectral features.

    Nonetheless, we discarded any spectrum with a global signal-to-noise ratio below 1.0 to ensure a minimum level of reliability in the fitting process.

    Motivated by the inconsistent performance of the reduced chi-squared statistic on low-S/N Euclid spectra, we explored alternative similarity metrics for spectroscopic validation.
    We found that cosine similarity significantly outperforms the reduced chi-squared in suppressing false positives, by a factor of 3–4, particularly for low signal-to-noise spectra, which are prevalent in our initial color-based photometric selection due to the intrinsic faintness of the targeted objects.
    While in some cases the reduced chi-squared yields slightly better spectral fits, we adopted cosine similarity for the spectroscopic candidate selection due to its superior robustness in noisy regimes.
    The use of the reduced chi-squared as a comparison metric is retained for the final spectral classification step.

    Cosine similarity has been successfully applied in other spectral classification contexts, for instance in~\cite{Yang+2024} or~\cite{Viscasillas+2024}.
    It measures the cosine of the angle between two non-zero vectors—here, the observed and template spectra.
    It is defined as the L2-normalized dot product:
    \begin{equation}
        k(A, B) = \frac{\langle A, B \rangle}{\|A\| \cdot \|B\|} = \frac{\sum_{i=1}^{n} A_i B_i}{\sqrt{\sum_{i=1}^{n} A_i^2} \cdot \sqrt{\sum_{i=1}^{n} B_i^2}}\label{eq:cosine-similarity}
    \end{equation}
    where $A$ and $B$ denote the flux values of the observed and template spectra, respectively.
    A value close to 1 indicates a strong match, corresponding to a small angle between the spectra in vector space.

    Each Euclid spectrum was compared to the full grid of spectral templates using the cosine similarity metric.
    Sources whose best-matching template corresponded to a spectral type between T0 and T9 were flagged as potential candidates.
    This resulted in an initial selection of 453 sources for further analysis.

    However, only 15 of these objects were ultimately classified as high-confidence brown dwarf candidates.
    This significant reduction is primarily due to the limitations of the Euclid NISP instrument and the low signal-to-noise of many extracted spectra.
    NISP is designed for broad cosmological objectives, such as galaxy redshift measurements and weak lensing studies.
    Its resolving power of R $>$ 480 (red grisms) for an object of 0\farcs5 diameter is well suited for detecting broad emission features in distant galaxies, but not optimized for the identification of faint point-like objects such as late-type brown dwarfs.

    Moreover, the spectral quality for many sources in the Q1 release is modest, especially for fainter objects.
    The combination of low resolution, spectral artifacts, and noise often result in ambiguous or poorly constrained matches, even when the photometry is consistent with ultracool spectral types.

    To address this, we performed a second-stage visual inspection of all fitted spectra and their associated template overlays.
    This step helped us eliminate sources with insufficient or poor signal in key diagnostic regions (e.g., CH$_4$ absorption at 1.3 and \SI{1.6}{\micro\metre}).

    This two-stage validation procedure was performed using \texttt{Euclid\_tools} \citep{Kiwy+2025}, a Python package we developed to facilitate the analysis of Euclid data.
    The package provides tools for spectral extraction, template matching, and visual inspection, streamlining the process of identifying ultracool dwarf candidates in large datasets like Euclid Q1.

    After this rigorous vetting, only 15 sources were retained as robust, high-confidence mid-to-late T dwarf candidates.
    Eight of these sources are likely new discoveries, while the remaining seven have been previously reported in the literature (Table~\ref{tab:object-table}).

    \begin{deluxetable*}{lllrrll}
        \tablecaption{Summary of the 15 high-confidence mid-to-late T dwarf candidates identified in this work \label{tab:object-table}}
        \tablehead{
            \colhead{Object Name} &
            \colhead{Short Name} &
            \colhead{Object ID} &
            \colhead{RA} &
            \colhead{Dec} &
            \colhead{Lit. Ref.} &
            \colhead{Lit. SpT} \\
            \colhead{~} &
            \colhead{~} &
            \colhead{~} &
            \colhead{(deg)} &
            \colhead{(deg)} &
            \colhead{~} &
            \colhead{~}
        }
        \startdata
        E032252.96\ms265010.92 & E0322\ms2650 & \ms507206654268363669 & 50.720665  & \ms26.836367 &         &            \\
        E032759.32\ms274940.51 & E0328\ms2749 & \ms519971822278279190 & 51.997182  & \ms27.827919 & 2, 3    & T6, T6     \\
        E032925.80\ms290153.42 & E0329\ms2901 & \ms523574860290315045 & 52.357486  & \ms29.031505 & 2       & T4         \\
        E033117.79\ms261858.95 & E0331\ms2618 & \ms528241075263163744 & 52.824107  & \ms26.316374 & 2       & T5         \\
        E033118.45\ms272549.34 & E0331\ms2725 & \ms528268759274303728 & 52.826876  & \ms27.430373 &         &            \\
        E033126.66\ms271240.56 & E0331\ms2712 & \ms528610966272112669 & 52.861097  & \ms27.211267 &         &            \\
        E034825.35\ms501644.69 & E0348\ms5016 & \ms571056342502790814 & 57.105634  & \ms50.279081 & 3       & T          \\
        E035231.98\ms491058.81 & E0352\ms4910 & \ms581332495491830038 & 58.133249  & \ms49.183004 & 1, 2    & T7, T7     \\
        E035909.93\ms474057.42 & E0359\ms4740 & \ms597913643476826162 & 59.791364  & \ms47.682616 & 1, 2, 3 & T8, T7, T7 \\
        E040322.22\ms472516.51 & E0403\ms4725 & \ms608425655474212533 & 60.842566  & \ms47.421253 &         &            \\
        E040417.08\ms463943.06 & E0404\ms4639 & \ms610711868466619603 & 61.071187  & \ms46.661960 &         &            \\
        E041818.40\ms480412.44 & E0418\ms4804 & \ms645766530480701214 & 64.576653  & \ms48.070121 &         &            \\
        E174556.40\ps645937.11 & E1745\ps6459 & 2664850113649936423 & 266.485011 & 64.993642  & 1, 2, 3, 4 & T7, T6, T6, T7 \\
        E174710.98\ps645429.83 & E1747\ps6454 & 2667957458649082863 & 266.795746 & 64.908286  &         &            \\
        E180505.04\ps644407.85 & E1805\ps6444 & 2712709955647355135 & 271.270996 & 64.735514  &         &            \\
        \enddata
        \tablecomments{
            The table lists the assigned object name, object short name, Euclid object identifier, equatorial coordinates (RA, Dec; J2000), and relevant literature references. Spectral types from the literature are indicated in the same order as the references. The spectral classifications in \citet{Zerjal+2025} are those reported by \citet{Dominguez-Tagle+2025}. All candidates were selected based on near-infrared color criteria, spectral template fitting, and visual inspection of their Euclid NISP spectra.
        }
        \tablerefs{
            (1) \citet{JYZhang+2024}; (2) \citet{Dominguez-Tagle+2025}; (3) \citet{Zerjal+2025}; (4) \citet{Mace+2013}
        }
    \end{deluxetable*}

    \section{Candidate Analysis}\label{sec:candidate-analysis}

    \subsection{Morphological and Photometric Properties}\label{subsec:morphological-and-photometric-properties}

    To evaluate the morphological properties and photometric reliability of the candidate sample, we examined several diagnostic parameters available in the Euclid Q1 merged catalogue.
    Four classification-related fields—\texttt{blended\_prob}, \texttt{variable\_flag}, \texttt{binary\_flag}, and \texttt{extended\_prob}—are missing for all sources in our final sample and thus could not be used for quality assessment.

    The \texttt{point\_like\_prob} parameter, which quantifies the likelihood that a source is point-like, is only available for 10 of the 15 candidates.
    However, its values are consistently low ($<$ 0.5), suggesting non-stellar morphologies.
    This is likely an artifact of the low signal-to-noise ratios and NIR-only detection for faint objects such as mid-to-late T dwarfs.
    In several cases, visual inspection of the image cutouts reveals clearly point-like profiles despite the low \texttt{point\_like\_prob}, indicating that this parameter is not fully reliable for faint brown dwarf candidates.

    As a more robust alternative, we considered the \texttt{mumax\_minus\_mag} parameter, which is defined as the difference between the peak surface brightness (\texttt{mu\_max}) and the total magnitude (\texttt{mag\_stargal\_sep}). This quantity, expressed in mag\,arcsec$^{-2}$, serves as a proxy for the compactness of the source: larger negative values typically correspond to more centrally concentrated (i.e., point-like) objects.
    All candidates have valid entries for this parameter, and the values support the compact nature expected for brown dwarfs (Table~\ref{tab:morphology-table}).

    The \texttt{fwhm} parameter provides the full width at half maximum of the source, as used by \texttt{a-phot} to define photometric apertures.
    All candidates show \texttt{fwhm} values between 1\farcs17 and 1\farcs39, reinforcing their point-like character.

    Lastly, the \texttt{ellipticity} parameter quantifies the elongation of the source based on its minor-to-major axis ratio.
    The candidates show low ellipticity values overall, further supporting their classification as substellar sources.

    Together, these parameters help establish the morphological integrity of the candidates and provide supporting evidence for their compact, point-like nature—characteristics consistent with the expectations for distant brown dwarfs.
    \texttt{mumax\_minus\_mag}, \texttt{fwhm}, and \texttt{ellipticity} have been successfully used by \citet{Zerjal+2025} to isolate point-like objects in their photometric search for ultracool dwarfs in the Euclid Q1 dataset.

    Consistent with their compact morphology, the candidates exhibit faint near-infrared fluxes, with $J_E$-band magnitudes ranging from 19.44 to 21.95 mag, a median of 20.67 mag, and a mean of 20.80 mag—values consistent with expectations for cool, distant brown dwarfs (Table~\ref{tab:photometry-table}).
    Photometric uncertainties are low across all near-infrared bands, with median values of 0.017 mag in $Y_E$, 0.014 mag in $J_E$, and 0.015 mag in $H_E$, enabling reliable color measurements.

    $VIS$-band magnitudes, where available, range from 24.47 to 26.75 mag (median 25.56 mag), and the corresponding flux-based signal-to-noise ratio varies between 8.04 and 35.42, with a median of 21.08.
    This highlights the suppressed optical flux of T dwarfs and supports our decision to exclude VIS photometry from the color-based selection.

    In contrast, the near-infrared bands show significantly higher signal-to-noise ratios.
    Median flux S/N values are 63.25 in $Y_E$, 74.90 in $J_E$, and 74.05 in $H_E$, with mean values of 57.37, 76.91, and 68.20, respectively.
    These consistently high S/N values across the near-infrared support the robustness of both the photometric colors and the extracted spectra.

    Color indices further validate the T dwarf classification.
    The $Y_E - J_E$ color spans 0.219 to 0.571 mag (median 0.403 mag), while $J_E - H_E$ values range from –0.484 to –0.106 mag (median –0.303 mag), reflecting the strong methane absorption characteristic of T dwarf atmospheres.

    Table~\ref{tab:statistical-summary} summarizes the properties of the 15 high-confidence T dwarf candidates in our final sample.
    Their photometric and morphological characteristics are consistent with expectations for mid-to-late T spectral types.

    \begin{deluxetable}{lccccc}
        \tabletypesize{\footnotesize}
        \tablecaption{Morphological parameters of the 15 mid-to-late T dwarf candidates \label{tab:morphology-table}}
        \tablehead{
            \colhead{Object Name} &
            \colhead{$P_\mathrm{point}$} &
            \colhead{$P_\mathrm{spur}$} &
            \colhead{$\mu_{\max} - m$} &
            \colhead{FWHM} &
            \colhead{Ellipticity} \\
            \colhead{~} &
            \colhead{~} &
            \colhead{~} &
            \colhead{($\mathrm{mag\,arcsec^{-2}}$)} &
            \colhead{($\arcsec$)} &
            \colhead{~}
        }
        \startdata
        E0322\ms2650 & 0.00   & 0.03 & \ms1.94 & 1.27 & 0.45 \\
        E0328\ms2749 & 0.45   & 0.07 & \ms2.85 & 1.29 & 0.04 \\
        E0329\ms2901 & \ldots & 0.03 & \ms1.23 & 1.35 & 0.05 \\
        E0331\ms2618 & 0.04   & 0.08 & \ms2.97 & 1.17 & 0.13 \\
        E0331\ms2725 & \ldots & 0.02 & \ms1.25 & 1.39 & 0.33 \\
        E0331\ms2712 & \ldots & 0.02 & \ms1.35 & 1.25 & 0.09 \\
        E0348\ms5016 & 0.32   & 0.04 & \ms2.83 & 1.29 & 0.15 \\
        E0352\ms4910 & 0.18   & 0.07 & \ms2.61 & 1.24 & 0.19 \\
        E0359\ms4740 & 0.50   & 0.07 & \ms2.90 & 1.25 & 0.05 \\
        E0403\ms4725 & 0.00   & 0.03 & \ms2.34 & 1.26 & 0.15 \\
        E0404\ms4639 & \ldots & 0.02 & \ms1.34 & 1.25 & 0.04 \\
        E0418\ms4804 & \ldots & 0.08 & \ms1.06 & 1.23 & 0.05 \\
        E1745\ps6459 & 0.33   & 0.13 & \ms2.75 & 1.28 & 0.03 \\
        E1747\ps6454 & 0.10   & 0.04 & \ms2.51 & 1.24 & 0.49 \\
        E1805\ps6444 & 0.14   & 0.04 & \ms2.63 & 1.25 & 0.21 \\
        \enddata
        \tablecomments{
            Columns list the object name, point-like probability ($P_\mathrm{point}$), spurious probability ($P_\mathrm{spur}$), difference between peak surface brightness and total magnitude ($\mu_{\max} - m$) as a proxy for compactness, full width at half maximum (FWHM) in arcseconds, and ellipticity. These parameters help assess the point-like nature and reliability of the detections. Missing values are indicated by an ellipsis.
        }
    \end{deluxetable}

    \begin{deluxetable*}{lrrrrrrrrrrrr}
        \tablecaption{Photometric properties of the 15 mid-to-late T dwarf candidates \label{tab:photometry-table}}
        \tablehead{
            \colhead{Object Name} &
            \colhead{$VIS$} &
            \colhead{$\sigma_{VIS}$} &
            \colhead{$Y_E$} &
            \colhead{$\sigma_{Y}$} &
            \colhead{$J_E$} &
            \colhead{$\sigma_{J}$} &
            \colhead{$H_E$} &
            \colhead{$\sigma_{H}$} &
            \colhead{$Y_E - J_E$} &
            \colhead{$\sigma_{Y - J}$} &
            \colhead{$J_E - H_E$} &
            \colhead{$\sigma_{J - H}$}\\
            \colhead{~} &
            \colhead{(mag)} &
            \colhead{(mag)} &
            \colhead{(mag)} &
            \colhead{(mag)} &
            \colhead{(mag)} &
            \colhead{(mag)} &
            \colhead{(mag)} &
            \colhead{(mag)} &
            \colhead{(mag)} &
            \colhead{(mag)} &
            \colhead{(mag)} &
            \colhead{(mag)}
        }
        \startdata
        E0322\ms2650 & 26.750 & 0.135  & 21.661 & 0.028 & 21.297 & 0.021 & 21.403 & 0.020 & 0.364 & 0.035 & \ms0.106 & 0.029 \\
        E0328\ms2749 & 24.806 & 0.040  & 19.916 & 0.013 & 19.436 & 0.008 & 19.739 & 0.011 & 0.480 & 0.015 & \ms0.303 & 0.014 \\
        E0329\ms2901 & \ldots & \ldots & 21.179 & 0.017 & 20.670 & 0.014 & 20.882 & 0.015 & 0.509 & 0.022 & \ms0.212 & 0.021 \\
        E0331\ms2618 & 26.207 & 0.112  & 21.957 & 0.029 & 21.399 & 0.020 & 21.851 & 0.026 & 0.558 & 0.035 & \ms0.452 & 0.033 \\
        E0331\ms2725 & \ldots & \ldots & 22.428 & 0.043 & 21.917 & 0.032 & 22.321 & 0.043 & 0.511 & 0.054 & \ms0.404 & 0.054 \\
        E0331\ms2712 & \ldots & \ldots & 22.055 & 0.039 & 21.836 & 0.027 & 22.105 & 0.035 & 0.219 & 0.047 & \ms0.269 & 0.044 \\
        E0348\ms5016 & 25.493 & 0.048  & 20.863 & 0.017 & 20.497 & 0.014 & 20.660 & 0.014 & 0.366 & 0.022 & \ms0.163 & 0.020 \\
        E0352\ms4910 & 24.468 & 0.033  & 19.853 & 0.010 & 19.450 & 0.008 & 19.934 & 0.008 & 0.403 & 0.013 & \ms0.484 & 0.011 \\
        E0359\ms4740 & 24.807 & 0.031  & 20.091 & 0.012 & 19.615 & 0.007 & 20.002 & 0.009 & 0.476 & 0.014 & \ms0.387 & 0.011 \\
        E0403\ms4725 & 26.396 & 0.115  & 21.859 & 0.030 & 21.530 & 0.025 & 21.907 & 0.030 & 0.329 & 0.039 & \ms0.377 & 0.039 \\
        E0404\ms4639 & \ldots & \ldots & 22.009 & 0.033 & 21.684 & 0.025 & 21.895 & 0.034 & 0.325 & 0.041 & \ms0.211 & 0.042 \\
        E0418\ms4804 & \ldots & \ldots & 22.524 & 0.042 & 21.953 & 0.029 & 22.290 & 0.039 & 0.571 & 0.051 & \ms0.337 & 0.049 \\
        E1745\ps6459 & 24.894 & 0.037  & 20.283 & 0.014 & 19.846 & 0.013 & 20.272 & 0.013 & 0.437 & 0.019 & \ms0.426 & 0.018 \\
        E1747\ps6454 & 25.732 & 0.056  & 20.636 & 0.015 & 20.316 & 0.011 & 20.549 & 0.012 & 0.320 & 0.019 & \ms0.233 & 0.016 \\
        E1805\ps6444 & 25.634 & 0.085  & 21.006 & 0.014 & 20.620 & 0.011 & 20.731 & 0.012 & 0.386 & 0.018 & \ms0.111 & 0.016 \\
        \enddata
        \tablecomments{
            Columns list the object name, apparent magnitudes ($VIS$, $Y_E$, $J_E$, $H_E$) and their associated uncertainties ($\sigma_{VIS}$, $\sigma_Y$, $\sigma_J$, $\sigma_H$) in the VIS and NISP bands. The table also includes the $Y_E - J_E$ and $J_E - H_E$ colour indices, along with their uncertainties ($\sigma_{Y - J}$, $\sigma_{J - H}$), which are key diagnostics for T dwarf classification. Magnitudes are derived from template-fitting photometry in the NIR bands and PSF-fitting photometry in the $VIS$ band, and are given in the AB system. Missing VIS photometry is indicated by an ellipsis.
        }
    \end{deluxetable*}

    \begin{deluxetable}{lrrrr}
        \tabletypesize{\footnotesize}
        \tablecaption{Statistical summary of the 15 mid-to-late T dwarf candidates \label{tab:statistical-summary}}
        \tablehead{
            \colhead{Parameter} &
            \colhead{Min} &
            \colhead{Max} &
            \colhead{Median} &
            \colhead{Mean}
        }
        \startdata
        $VIS$ (mag)                           & 24.468  & 26.750   & 25.563   & 25.519   \\
        $\sigma_{\mathrm{VIS}}$ (mag)         & 0.031   & 0.135    & 0.052    & 0.069    \\
        $Y_E$ (mag)                           & 19.853  & 22.524   & 21.179   & 21.221   \\
        $\sigma_{Y}$ (mag)                    & 0.010   & 0.043    & 0.017    & 0.024    \\
        $J_H$ (mag)                           & 19.436  & 21.953   & 20.670   & 20.804   \\
        $\sigma_{J}$ (mag)                    & 0.007   & 0.032    & 0.014    & 0.018    \\
        $H_E$ (mag)                           & 19.739  & 22.321   & 20.882   & 21.103   \\
        $\sigma_{H}$ (mag)                    & 0.008   & 0.043    & 0.015    & 0.021    \\
        $Y_E - J_E$ (mag)                     & 0.219   & 0.571    & 0.403    & 0.417    \\
        $J_E - H_E$ (mag)                     & \ms0.484  & \ms0.106   & \ms0.303   & \ms0.298   \\
        S/N$_{\mathrm{flux}}~VIS$            & 8.040   & 35.420   & 21.075   & 20.800   \\
        S/N$_{\mathrm{flux}}~Y_E$            & 25.190  & 103.660  & 63.250   & 57.372   \\
        S/N$_{\mathrm{flux}}~J_E$            & 34.340  & 153.050  & 74.900   & 76.915   \\
        S/N$_{\mathrm{flux}}~H_E$            & 25.000  & 137.200  & 74.050   & 68.204   \\
        P$_{\mathrm{point}}$                  & 0.000   & 0.500    & 0.160    & 0.206    \\
        P$_{\mathrm{spur}}$                   & 0.020   & 0.130    & 0.040    & 0.051    \\
        $\mu_{\max} - m$ (mag\,arcsec$^{-2}$) & \ms2.970  & \ms1.060   & \ms2.510   & \ms2.171   \\
        FWHM (arcsec)                         & 1.170   & 1.390    & 1.250    & 1.267    \\
        Ellipticity                           & 0.030   & 0.490    & 0.130    & 0.163    \\
        S/N$_{\mathrm{spec}}$                 & 1.169   & 7.611    & 3.372    & 3.512    \\
        SpT$_{\mathrm{B}}$                    & 2.000   & 7.000    & 4.000    & 4.067    \\
        SpT$_{\mathrm{T}}$                    & 2.000   & 7.000    & 5.000    & 4.667    \\
        \mean~\dphot~(pc)      & 30.100  & 184.051  & 105.288  & 100.339  \\
        \median~\dphot~(pc)    & 30.269  & 185.439  & 99.174   & 99.440   \\
        \std~\dphot~(pc)        & 0.643   & 11.601   & 3.088    & 5.106    \\
        \enddata
        \tablecomments{
            Magnitudes are in the AB system, photometric uncertainties are denoted by $\sigma$, $Y_E - J_E$ and $J_E - H_E$ are color indices, and S/N values are reported from fluxes. P$_{\mathrm{point}}$ and P$_{\mathrm{spur}}$ are the probabilities of being point-like or spurious detections, respectively. $\mu_{\max} - m$ is the difference between peak surface brightness and total magnitude. S/N$_{\mathrm{spec}}$ is the S/N from the Euclid spectra. SpT$_{\mathrm{B}}$ and SpT$_{\mathrm{T}}$ are numerical spectral types inferred from Burgasser and Theissen templates, respectively. $d_{\mathrm{phot}}$ values are the mean, median, and standard deviation of the photometric distance estimates derived from $Y_E$, $J_E$, and $H_E$ bands.
        }
    \end{deluxetable}

    \subsection{Spectral Classification via Template Matching}\label{subsec:spectral-classification-via-template-matching}

    To determine the spectral types of our candidate brown dwarfs, we compared their Euclid NISP spectra to two sets of empirical templates: (1) M, L, and T dwarf spectra from~\citealt{Burgasser+2017} (used in Section \ref{sec:spectroscopic-validation}, hereafter: Burgasser templates), and (2) L, T, and Y dwarf spectra from \citet{Theissen+2022} (hereafter: Theissen templates).
    Each Euclid spectrum was trimmed to the 1.22–\SI{1.88}{\micro\metre} range to exclude low-S/N edges and then smoothed using a Savitzky–Golay filter with a window length of 11 and a polynomial order of 2, which reduces pixel-scale noise while preserving broad spectral features.
    Zero-flux values were masked to prevent spurious contributions to the chi-squared statistic during template fitting.

    The template spectra were processed similarly: they were trimmed to the same wavelength range and interpolated onto the Euclid wavelength grid.
    Each template was then flux-scaled to match the Euclid spectrum by applying a normalization factor equal to the ratio of their mean fluxes over the trimmed interval.
    For each candidate, we computed reduced chi-squared ($\chi^2_\nu$) values across all templates in both libraries and identified the best-fitting template and its associated spectral type.
    $\chi^2_\nu$ as employed here is defined as:
    \begin{equation}
        \chi^2_\nu = \frac{1}{\nu} \sum_{i} \frac{\left(F_{\mathrm{o},i} - F_{\mathrm{t},i}\right)^2}{\sigma_{\mathrm{o},i}^2}\label{eq:chi-square}
    \end{equation}
    where $F_{\mathrm{o},i}$ and $F_{\mathrm{t},i}$ are the observed and scaled template fluxes at pixel $i$, $\sigma_{\mathrm{o},i}$ is the observed flux uncertainty, and $\nu$ is the number of degrees of freedom (number of data points minus number of fitted parameters).

    Figures~\ref{fig:comparison-with-burgasser-templates} and~\ref{fig:comparison-with-theissen-templates} present the best-fitting empirical templates for each candidate, together with their spectral types, and highlight key molecular absorption features including CH$_4$, H$_2$O, K\,$\textsc{i}$, and FeH.

    \subsection{Spectral Evaluation of Candidates}\label{subsec:spectral-evaluation-of-candidates}

    T dwarfs are characterized by distinctive near-infrared spectral features dominated by strong molecular absorption bands, particularly from methane (CH$_4$) and water (H$_2$O). The presence of deep CH$_4$ absorption troughs near \SI{1.3}{\micro\metre}, \SI{1.6}{\micro\metre}, and \SI{2.2}{\micro\metre} marks the transition from late-L to T spectral types~\citep{Burgasser+2002}.
    Within the Euclid NISP spectral wavelength range (1.2–\SI{1.9}{\micro\metre}, red grisms), the most prominent diagnostics are the suppressed $H$-band flux due to methane absorption and a peaked $J$-band, with minimal flux in between.
    These features enable spectral classification even at low resolution, provided the signal-to-noise ratio is sufficient.
    However, some brown dwarf candidates lie near the detection limit of the instrument, leading to noisy or incomplete spectra.
    In addition, variations in metallicity may lead to imperfect template fits, since the standard templates do not cover a range of metallicities for each spectral type.
    These limitations must be considered when assessing spectral fits.

    In Appendix~\ref{sec:spectral-characteristics-of-candidates}, we present detailed evaluations of the spectra of our 15 candidate brown dwarfs, highlighting the characteristic near-infrared features used for their classification and noting cases where low signal-to-noise affects the spectral assessment.

    \subsection{Photometric Type and Distance Estimations}\label{subsec:photometric-type-and-distance-estimations}

    To complement the spectral classification, we estimated a photometric spectral type for each candidate based on its Euclid near-infrared colors.
    Using empirical polynomial relations published by \citet{Sanghi+2024}, we computed synthetic absolute magnitudes in the Euclid NISP $Y_E$, $J_E$, and $H_E$ bands, separately for field-age and young objects.
    For each spectral type, we also derived color indices ($Y_E - J_E$, $Y_E - H_E$, and $J_E - H_E$), capturing characteristic trends across the M, L, and T dwarf sequence.
    Table~\ref{tab:abs-mag-table} lists the resulting absolute magnitudes and colors, which form the basis for our photometric type and distance estimates.

    A two-dimensional nearest-neighbor search was then performed in color–color space ($Y_E - J_E$ vs. $J_E - H_E$) using a $k$-nearest neighbor (k-NN) classifier with Euclidean distance as the metric.
    For each candidate, we identified the closest color match within the reference grid, along with its spectral type and associated fit statistic.
    To mitigate boundary effects, we excluded edge cases—specifically T9 classifications—whenever their nearest-neighbor distance exceeded two spectral subtypes.

    In several cases, noticeable discrepancies arise between the photometric spectral types and those derived from template matching (see Table~\ref{tab:photometric-distance-table}); for example, E0328\ms2749, E0359\ms4740, and E1745\ps6459 differ by up to four subtypes from the spectral types inferred using the Burgasser templates.
    This mismatch likely reflects the intrinsic color degeneracy of the late-T regime, where near-infrared colors do not follow a strictly linear trend with spectral type (see color–spectral type relations in Table~\ref{tab:abs-mag-table}).

    Because our distance estimates rely on spectral types from template fitting, we also examined the consistency between the two template libraries.
    The spectral types obtained from the Theissen templates are broadly consistent with those inferred from the Burgasser templates, typically agreeing within one subtype.
    We therefore adopted the mean of the two classifications as the final spectral type for each source, which then served as the basis for photometric distance estimation.

    Using these adopted spectral types, we inferred absolute magnitudes in the Euclid $Y_E$, $J_E$, and $H_E$ bands from the empirical relations of \citet{Sanghi+2024}.
    From these and the observed apparent magnitudes, we computed individual band-specific photometric distances via the distance modulus (Table~\ref{tab:photometric-distance-table}).
    We report both the mean and median of the three distances as the final estimate, while the dispersion among them was used as an estimate of the uncertainty.
    This approach provides consistent distance estimates in the absence of parallaxes while accounting for band-to-band variation and photometric noise.

    \begin{deluxetable}{lllllrrr}
        \tabletypesize{\footnotesize}
        \tablecaption{Photometric and spectral types with photometric distance estimates for our 15 candidates \label{tab:photometric-distance-table}}
        \tablehead{
            \colhead{~} &
            \multicolumn{4}{c|}{Type} &
            \multicolumn{3}{c}{\dphot} \\
            \cline{2-5} \cline{6-8}
            \colhead{Object Name} &
            \colhead{PHO} &
            \colhead{S$_B$} &
            \colhead{S$_T$} &
            \colhead{ADP} &
            \colhead{\mean} &
            \colhead{\median} &
            \colhead{\std} \\
            \colhead{~} &
            \colhead{~} &
            \colhead{~} &
            \colhead{~} &
            \colhead{~} &
            \colhead{(pc)} &
            \colhead{(pc)} &
            \colhead{(pc)}
        }
        \startdata
        E0322\ms2650 & \ldots & T7 & T7 & T7 & 66  & 68  & 5  \\
        E0328\ms2749 & T8 field   & T4 & T5 & T5 & 51  & 51  & 2  \\
        E0329\ms2901 & T4 young   & T2 & T2 & T2 & 112 & 107 & 11 \\
        E0331\ms2618 & T5 young   & T4 & T5 & T5 & 130 & 134 & 7  \\
        E0331\ms2725 & T4 young   & T4 & T4 & T4 & 184 & 185 & 12 \\
        E0331\ms2712 & T5 field   & T4 & T5 & T5 & 147 & 147 & 3  \\
        E0348\ms5016 & \ldots & T3 & T4 & T4 & 90  & 90  & 2  \\
        E0352\ms4910 & T7 field   & T6 & T7 & T7 & 30  & 30  & 1  \\
        E0359\ms4740 & T8 field   & T4 & T5 & T5 & 56  & 57  & 2  \\
        E0403\ms4725 & T8 field   & T5 & T6 & T6 & 107 & 107 & 1  \\
        E0404\ms4639 & T5 field   & T4 & T5 & T5 & 138 & 137 & 1  \\
        E0418\ms4804 & T4 young   & T6 & T6 & T6 & 134 & 128 & 9  \\
        E1745\ps6459 & T8 field   & T4 & T5 & T5 & 62  & 63  & 2  \\
        E1747\ps6454 & T5 field   & T2 & T2 & T2 & 93  & 86  & 11 \\
        E1805\ps6444 & \ldots & T2 & T2 & T2 & 105 & 99  & 9  \\
        \enddata
        \tablecomments{
            The table lists the photometric spectral type (PHO) inferred from near-infrared colors, the spectral types derived from Burgasser (S$_B$) and Theissen (S$_T$) templates, the adopted spectral type (ADP; mean of S$_B$ and S$_T$) used for distance estimation, and the mean, median, and standard deviation of the photometric distances (in parsecs). Photometric types are not provided for candidates with unreliable classifications (e.g., edge cases such as T9). Objects labeled as "young" reflect the outcome of the nearest-neighbour classification and should not be interpreted as evidence of confirmed youth.
        }
    \end{deluxetable}

    \subsection{Visual Inspection of Candidates}\label{subsec:visual-inspection-of-candidates}

\begin{figure*}
        \begin{center}
            \includegraphics[width=15cm]{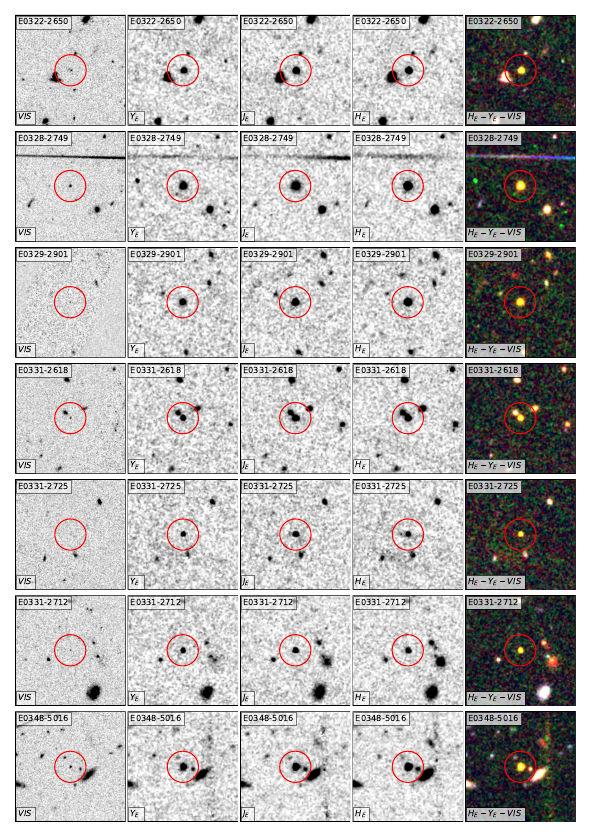}
        \end{center}
        \caption{
            Euclid $20\arcsec \times 20\arcsec$ cutouts of the 15 brown dwarf candidates in four bands $VIS$, $Y_E$, $J_E$, $H_E$, and a color composite (left to right). Each candidate is centered within the red circle. The VIS cutouts illustrate that most candidates are barely detected at optical wavelengths, as expected for mid-to-late T dwarfs. In contrast, the majority are clearly visible in the $Y_E$, $J_E$ and $H_E$ bands. Although most candidates appear compact and isolated, some might be slightly blended by nearby background sources. All images are oriented with North up and East to the left.
        }
        \label{fig:cutouts}
    \end{figure*}

    Each of the 15 brown dwarf candidates was visually inspected using $20\arcsec \times 20\arcsec$ Euclid cutouts (Figure~\ref{fig:cutouts}) in the $VIS$, $Y_E$, $J_E$, and $H_E$ bands, together with a color composite where $H_E$, $Y_E$, and $VIS$ are mapped to the red, green, and blue channels, respectively.
    All candidates are clearly detected in the near-infrared $Y_E$, $J_E$, and $H_E$ bands, while $VIS$-band detections are generally faint, with some sources barely discernible—consistent with the intrinsically low optical fluxes of mid-to-late T dwarfs.
    Most objects appear compact and symmetric, with no significant signs of elongation or blending.
    A few lie in moderately crowded fields, where faint neighboring sources could influence aperture photometry or contaminate the extracted spectra.
    Overall, the majority of candidates exhibit isolated, point-like morphologies, supporting the reliability of both their photometric and spectroscopic measurements.

    In the following, we highlight noteworthy cases, focusing on potential blending or crowding issues that may affect the robustness of the photometry or spectra:

    \textbf{E0322\ms2650} Two overlapping bright sources (a star and a galaxy) are located to the left of the candidate, but they are sufficiently distant to avoid contaminating its photometry or spectrum.

    \textbf{E0331\ms2618} A nearby source lies approximately 1$\arcsec$ to the upper left, potentially introducing slight contamination in the photometry or spectrum.

    \textbf{E0348\ms5016} Two faint neighboring sources are visible within $\sim2\arcsec$ (top left and right), but are unlikely to affect the flux due to their faintness.
    A brighter galaxy lies $\sim4\arcsec$ to the bottom right, but its separation suggests negligible impact on the candidate's photometry.

    \textbf{E0352\ms4910} The candidate nearly overlaps two very faint background objects, whose contribution to the total flux is likely minimal.
    Despite this, the spectrum is the cleanest in the sample and supports a high-confidence T6 classification, though a T7 match is also plausible depending on the band.
    This minor discrepancy could be attributed to residual flux from the background sources.
    Two additional faint sources are located $\sim4\arcsec$ away (left and bottom left), but do not appear to affect the observed flux.

    \textbf{E1745\ps6459} A faint source is present within 1–2$\arcsec$ (top right), which may contribute slight contamination.
    However, the candidate’s spectrum appears clean and well matched, supporting a confident classification.

    \subsection{Detections in Other Surveys}\label{subsec:detections-in-other-surveys}

    We searched for detections of our candidates in relevant previous near- and mid-infrared surveys and found that eleven sources have valid photometry in either VHS or the VISTA Kilo-degree Infrared Galaxy Survey (VIKING; \citealt{Edge+2013}), including five detected only in the $J$-band.
    Six candidates also have mid-infrared counterparts in CatWISE2020~\citep{Marocco+2021}.
    The combined photometry from these surveys is reported in Table~\ref{tab:vhs_viking_wise_phot}, along with $W1 - W2$ colors and inferred photometric spectral types, which are broadly consistent with our adopted classifications.
    Magnitudes in this table are reported on the Vega system, in contrast to the AB system used for Euclid photometry throughout this paper.

    \begin{figure*}
        \begin{center}
            \includegraphics[width=15cm]{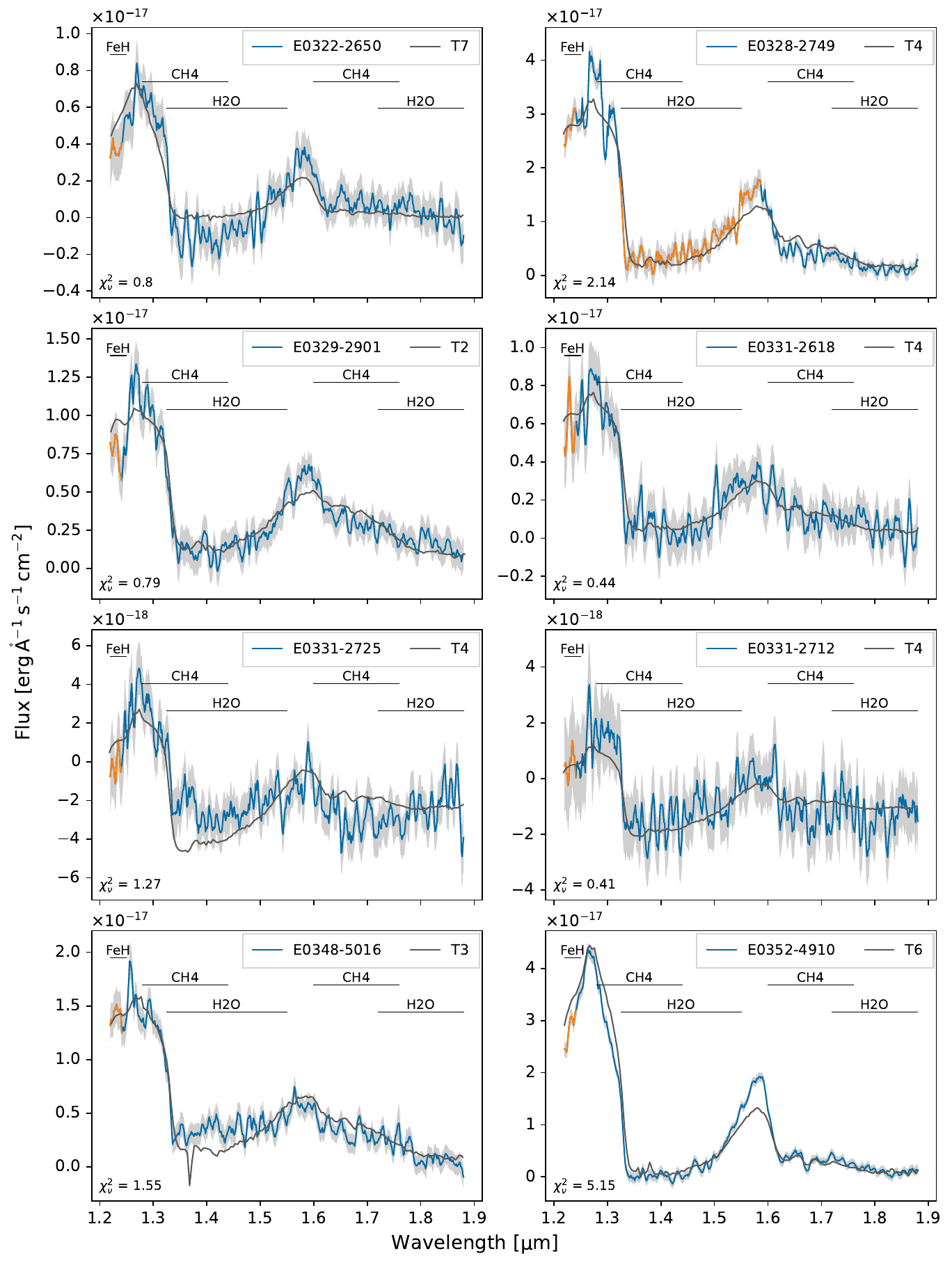}
        \end{center}
        \caption{
            Comparison of smoothed candidate spectra (blue curves) with the \textbf{Burgasser templates} (dark gray curves) for the 15 selected candidates. The orange curves represent flagged values (e.g., due to low quality or artifacts). The shaded gray area around each spectrum represents the flux uncertainty. Key molecular absorption features are indicated by horizontal black lines (FeH, CH$_4$, and H$_2$O). Specifically, CH$_4$ bands span 1.28 – \SI{1.44}{\micro\metre} in the $J$-band and 1.6 – \SI{1.76}{\micro\metre} in the $H$-band; H$_2$O bands span 1.325 - \SI{1.55}{\micro\metre} in the $J$-band and 1.72 - \SI{2.14}{\micro\metre} in the $H$-band. While several spectra exhibit clear methane absorption features characteristic of mid-to-late T dwarfs (e.g., at \SI{1.3}{\micro\metre}, \SI{1.6}{\micro\metre}), others show considerable noise or deviations from the templates, reflecting the limitations of low-resolution slitless spectroscopy and varying data quality in the Euclid Q1 release. The reduced chi-square fit statistic, displayed in the bottom-left corner, provides a quantitative measure of the fit quality.
        }
        \label{fig:comparison-with-burgasser-templates}
    \end{figure*}

    \begin{figure*}
        \begin{center}
            \includegraphics[width=15cm]{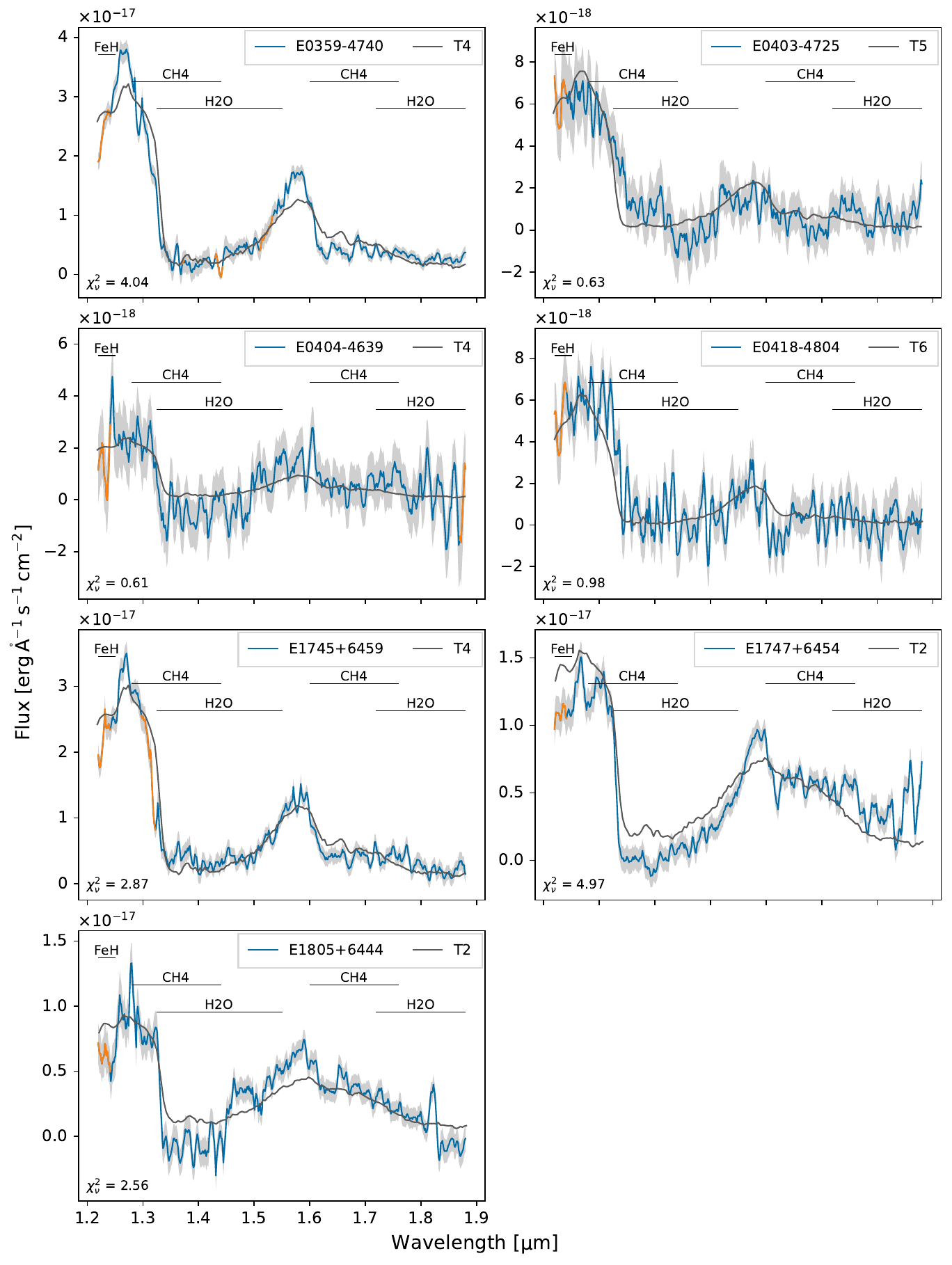}
        \end{center}
        \centering\textbf{Figure~\ref{fig:comparison-with-burgasser-templates}.} (Continued)
    \end{figure*}

    \begin{figure*}
        \begin{center}
            \includegraphics[width=15cm]{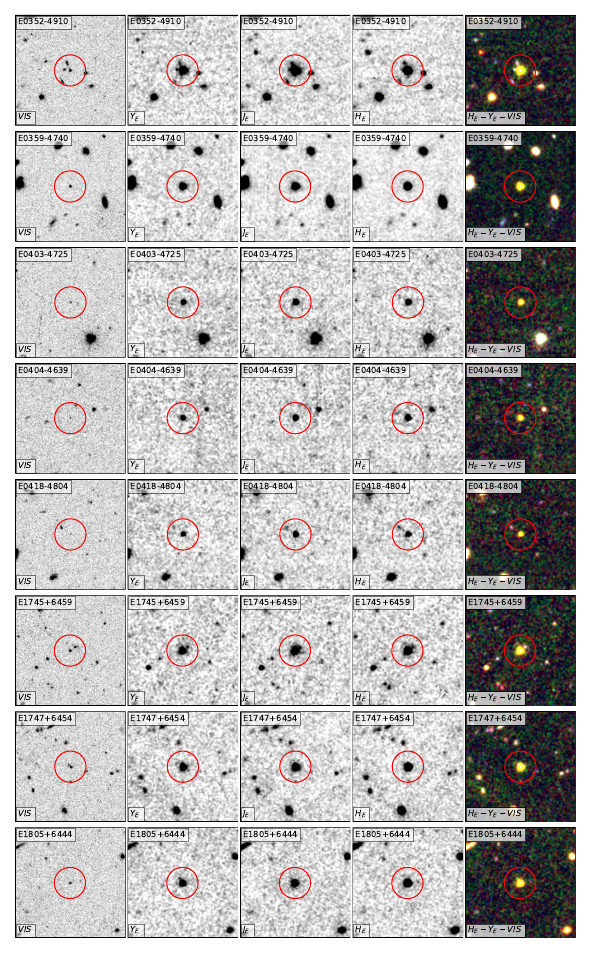}
        \end{center}
        \centering\textbf{Figure~\ref{fig:cutouts}.} (Continued)
    \end{figure*}

    \section{Refinement of Photometric Selection}\label{sec:refinement-of-photometric-selection}

    The initial candidate selection targeted mid-to-late T dwarfs in the Euclid Q1 merged catalogue by applying basic photometric constraints based on expected near-infrared colors.
    Specifically, we required reliable flux measurements in the $Y_E$, $J_E$, and $H_E$ bands, a magnitude cut of $<23$ mag in each band to ensure reasonable signal-to-noise, and characteristic color ranges of $0.1 < Y_E - J_E < 0.9$ and $J_E - H_E < -0.1$.
    This yielded a preliminary sample of 38,845 sources.

    To improve sample purity and reduce contamination, we applied additional criteria based on photometric and morphological quality indicators derived from our statistical analysis of confirmed candidates in Section~\ref{subsec:morphological-and-photometric-properties}.
    This included cuts on spurious detection probability ($<$ 0.2), ellipticity ($<$ 0.5), compactness index ($\mu_{\mathrm{\max}} - m$ between $-3.1$ and $-0.9$ mag\,arcsec$^{-2}$), and FWHM (between 1\farcs0 and 1\farcs5).
    Additionally, we imposed minimum signal-to-noise thresholds in each band: S/N$_Y > 24$, S/N$_J > 32$, and S/N$_H > 24$.
    After applying these refined constraints, the sample was reduced to a cleaner and more manageable set of 11,302 photometric candidates, while retaining all 15 spectroscopically identified mid-to-late T dwarfs.
    Table~\ref{tab:refined-selection-criteria} lists the complete morphological and photometric selection criteria applied to the Euclid Q1 merged catalogue.

    \begin{deluxetable}{ll}
        \tablecaption{Refined selection criteria for the Euclid Q1 merged catalogue \label{tab:refined-selection-criteria}}
        \tablehead{
            \colhead{Parameter} &
            \colhead{Criterion}
        }
        \startdata
        \texttt{det\_quality\_flag} & $= 0$                                        \\
        \texttt{spurious\_prob}     & $< 0.2$                                      \\
        \texttt{ellipticity}        & $< 0.5$                                      \\
        \texttt{mumax\_minus\_mag}  & $-3.1 \mathrm{~to~} -0.9$ mag\,arcsec$^{-2}$ \\
        \texttt{fwhm}               & $1\farcs0 \mathrm{~to~} 1\farcs5$        \\
        \texttt{flux\_y\_templfit}  & $> 0$                                        \\
        \texttt{flux\_j\_templfit}  & $> 0$                                        \\
        \texttt{flux\_h\_templfit}  & $> 0$                                        \\
        S/N $Y_E$                   & $> 24$                                       \\
        S/N $J_E$                   & $> 32$                                       \\
        S/N $H_E$                   & $> 24$                                       \\
        $Y_E, J_E, H_E$             & $< 23.0$ mag                                 \\
        $Y_E - J_E$                 & $0.1 \mathrm{~to~} 0.9$                      \\
        $J_E - H_E$                 & $< -0.1$                                     \\
        \enddata
        \tablecomments{
            Magnitudes are computed in the AB system as $m = -2.5 \log_{10}(f) + 23.9$, where $f$ is the flux in $\mu$Jy.
        }
    \end{deluxetable}

    \section{Discussion}\label{sec:discussion}

    The results presented in this work demonstrate the capability of the Euclid mission to identify and characterize brown dwarfs, despite its primary design for extragalactic cosmology.
    The combination of near-infrared imaging and slitless spectroscopy from the NISP instrument proves effective for detecting key features such as CH$_4$ and H$_2$O absorption, which are diagnostic of mid-to-late T dwarfs.

    \subsection{Comparison with Related Euclid-Based Studies}\label{subsec:comparison-with-related-euclid-based-studies}

    Our study complements recent efforts based on Euclid Q1 data, particularly the spectroscopic analysis of ultracool dwarfs (UCDs) by \citet{Dominguez-Tagle+2025}, the photometric selection conducted by \citet{Zerjal+2025}, and the spectroscopic work in the Euclid Deep Field North by \citet{Mohandasan+2025}.
    A key difference lies in the methodological design: \citet{Dominguez-Tagle+2025} perform a purely spectroscopic search, identifying and classifying UCDs directly from the NISP spectral archive, whereas \citet{Zerjal+2025} employ a purely photometric approach, relying on optical and near-infrared color cuts—most notably a $VIS - Y_E > 2.5$ criterion—to isolate high-purity UCD candidates.
    \citet{Mohandasan+2025}, in turn, use Euclid slitless spectroscopy to identify 33 new UCDs in the Deep Field North, primarily spanning spectral types M7 to T1 and $J_E$ magnitudes of 17–21.

    In contrast, our study adopts a hybrid approach that combines near-infrared color selection with subsequent spectroscopic validation, leveraging both photometric and spectroscopic information in an integrated framework.
    Unlike \citet{Zerjal+2025}, we do not require detections in the $VIS$-band, allowing us to probe a fainter and potentially more complete population of late-type brown dwarfs, including sources that are undetected in optical wavelengths.
    Our methodology is optimized for the recovery of high-confidence mid-to-late T dwarfs, whereas the aforementioned studies target a broader ultracool dwarf population.
    As such, our results serve as a complementary probe of the T dwarf population within the Euclid Q1 release, bridging the gap between the purely photometric and spectroscopic approaches adopted by other teams.

    \subsection{Spectral Classification and Uncertainties}\label{subsec:spectral-classification-and-uncertainties}

    Despite the use of two different template libraries (Burgasser and Theissen) enabling cross-validation of spectral classifications, the low spectral resolution and the relatively low S/N of many Euclid spectra impose limitations on the accuracy of subtype assignments, especially at the faint end.

    However, even for objects with adequate signal-to-noise, discrepancies persist.
    For instance, some of the nearest objects in our sample (e.g., E0328\ms2749, E0359\ms4740, E1745\ps6459) exhibit elevated fluxes in the J- and H-band peaks, while others (e.g., E0329\ms2901, E1747\ps6454, E1805\ps6444) show stronger FeH absorption than their best-fit templates.
    These differences are likely the result of genuine astrophysical diversity among cool brown dwarfs.
    Variations in metallicity (e.g., \citealt{McLean+2003, McLean+2007, Burgasser+2006, Burgasser+2025}), surface gravity (e.g., \citealt{Line+2015, Line+2017, Martin+2017}), or atmospheric chemistry and dynamics (e.g., \citealt{Cushing+2008, Burgasser+2010Clouds, Marley+2010, Suarez+2025}) can significantly alter the depth and shape of key absorption bands (e.g., H$_2$O and CH$_4$), as well as the overall near-IR continuum, leading to spectra that deviate from the solar-metallicity, field-gravity objects that define most existing templates.
    Additionally, unresolved multiplicity (e.g., L+T or T+Y systems) can produce blended spectral morphologies that mimic intermediate types or dilute characteristic features (e.g., \citealt{Burgasser+2007, Burgasser+2010Binaries}).
    Non-equilibrium chemistry (e.g., \citealt{Yamamura+2010, Zahnle+2014, Miles+2020, Turner+2025}), driven by vertical mixing, can further modify the relative strengths of CH$_4$ and CO bands, while atmospheric heterogeneities or rotationally modulated variability (e.g., \citealt{Buenzli+2012, Radigan+2012, Metchev+2015, Vos+2023, ChenXu+2024}) can lead to spectral shapes not well represented in static templates.
    These effects are compounded by the fact that the current T-dwarf standards are sparse, heterogeneous in origin, and do not fully sample the parameter space of gravity, metallicity, cloud properties, or atmospheric dynamics.
    As a result, some of our T dwarfs likely fall outside the empirical distribution spanned by the standards themselves, leading to imperfect or ambiguous template matches even when the observational data are of good quality.

    We also note that discrepancies between photometric and spectroscopic spectral types—sometimes spanning several subtypes—can arise from color degeneracies, spectral peculiarities, contamination from nearby sources, or differences in wavelength coverage (photometry: 950–2020 nm; spectroscopy, red grisms: 1206–1892 nm).
    Follow-up spectroscopy at higher resolution will be necessary to refine the classifications and investigate any outliers.

    While our classification procedure yields results broadly consistent with expectations from colors and spectral morphology, several candidates exhibit differences compared to previously published spectral types.
    For example, E0328\ms2749 is classified as T5 in our analysis versus T6 in \citet{Dominguez-Tagle+2025}; E0329\ms2901 is assigned T2 compared to a literature type of T4; E0359\ms4740 is classified as T5 versus T7 in the literature; and E1745\ps6459 is assigned T5 compared to T6 reported by \citet{Dominguez-Tagle+2025}.
    These differences are most likely the result of variations in the spectral classification methodology.
    Factors such as the choice of template library, the spectral similarity metric employed, the treatment of masked or flagged spectral regions, and the rules used to reconcile conflicting spectral types can all influence the final classification.
    Nonetheless, the offsets remain within 1–2 subtypes and do not affect the interpretation of these sources as mid-to-late T dwarfs.

    To assess the influence of flagged spectral regions on the classifications, we repeated the template fitting after removing all flagged pixels from the Euclid spectra.
    For all 15 candidates, including E0328\ms2749, which exhibits the largest contiguous region of flagged pixels (1.3–\SI{1.6}{\micro\metre}), the inferred spectral types remain unchanged.
    This indicates that, for the objects in our sample, the presence of flagged pixels does not influence the resulting classifications.
    We note that the impact of flagged regions will generally depend on the wavelength range and diagnostic features affected, and larger or strategically placed gaps could influence classifications in other cases.

    \subsection{Implications for Brown Dwarf Census}\label{subsec:implications-for-brown-dwarf-census}

    The identification of 15 high-confidence T dwarfs in just a fraction of the sky suggests that Euclid has strong potential to expand the census of substellar objects.
    The Euclid Q1 release covers 63.1~deg$^2$ of the Euclid Deep Fields, compared to the $\sim$14{,}500~deg$^2$ planned for the Euclid Wide Survey.
    Scaling our findings to this full area implies a potential yield of $\sim$3{,}450 mid-to-late T dwarf candidates using our methodology.
    This estimate indicates that Euclid has the potential to increase the known mid-to-late T dwarf population by several thousand objects, enabling stronger statistical constraints on the substellar mass function and improving our understanding of their spatial distribution and kinematics.
    This expanded census will provide valuable insights into the formation and evolution of substellar objects across diverse Galactic environments, and offer a critical foundation for future studies of atmospheric diversity and chemical composition in the coldest brown dwarfs.

    \subsection{Outlook and Future Work}\label{subsec:outlook-and-future-work}

    This study lays the groundwork for future brown dwarf searches in Euclid data releases.
    The forthcoming Data Release~1 (DR1), scheduled for October 2026, will cover approximately 1900~deg$^2$ of sky—about 30 times the area of Q1.
    Applying our methodology to this dataset could yield on the order of 450 additional mid-to-late T dwarfs.

    Follow-up spectroscopy of the current sample at higher signal-to-noise will be essential to refine spectral classifications, assess atmospheric properties, and confirm potential peculiarities such as low metallicity or unresolved binarity.

    \section{Conclusions}\label{sec:conclusions}

    We have conducted a targeted search for late-type brown dwarfs in the Euclid Q1 dataset by combining photometric selection with spectral template matching.
    Starting from an initial sample of 38,845 color-selected sources, we employed a two-stage validation process that included spectral comparison to empirical templates and visual inspection of key diagnostic features.
    This approach led to the identification of 15 high-confidence T dwarf candidates, spanning spectral types from T2 to T7.

    Each candidate was analyzed in detail, including assessments of morphological and photometric consistency, spectral fit quality, and the local imaging context to rule out contamination.
    Spectral types were assigned from high-confidence matches to two template libraries, with the mean value adopted as the final spectral type.

    Photometric distances were derived for each candidate using synthetic magnitudes from the \citet{Sanghi+2024} sample, yielding distances in the range of $\sim$30 to 185~pc.
    Eight of the 15 candidates in our sample appear to be previously unreported objects, whereas the remaining seven are consistent with known brown dwarfs in the literature.

    This study demonstrates that Euclid, despite being optimized for cosmology, can contribute significantly to the study of ultracool dwarfs.
    Our results highlight the potential of combining color selection and low-resolution NIR spectroscopy to robustly identify and characterize brown dwarfs in large survey data.
    Future data releases, with increased sky coverage and deeper sensitivity, are expected to substantially expand the known census of substellar objects.

    \section*{Acknowledgements}
    This work is based on observations made with the Euclid space telescope, a European Space Agency (ESA) mission with contributions from NASA and the Euclid Consortium.
    We acknowledge the use of data from the Euclid Quick Release 1 (Q1), publicly available through the ESA Euclid Science Archive \citep{Euclid+2025}.
    We thank the Euclid Consortium for their efforts in the design, construction, and data processing of the mission, and the ESA Science Data Centre (ESDC) for providing access to the Q1 merged catalogue and spectral products.
    We also recognize the Euclid Science Ground Segment (SGS) for their contributions to data processing and quality assessment, which enabled this analysis.
    This work makes use of data products from the VISTA Hemisphere Survey (VHS) and the VISTA Kilo-degree Infrared Galaxy (VIKING) Survey, obtained with the VIRCAM instrument on the Visible and Infrared Survey Telescope for Astronomy (VISTA), operated by the European Southern Observatory (ESO).
    The data were processed by the Cambridge Astronomical Survey Unit (CASU) and are publicly available through the VISTA Science Archive (VSA; \citealt{Cross+2012}), hosted by the Wide Field Astronomy Unit (WFAU) in Edinburgh.
    We acknowledge the use of data products from the Wide-field Infrared Survey Explorer, which is a joint project of the University of California, Los Angeles, and the Jet Propulsion Laboratory/California Institute of Technology, and NEOWISE which is a project of the Jet Propulsion Laboratory/California Institute of Technology.
    WISE and NEOWISE are funded by the National Aeronautics and Space Administration.
    This research has made use of the SIMBAD database~\citep{Wenger+2000}, the CDS VizieR catalogue access tool~\citep{Ochsenbein+2000}, and NASA's Astrophysics Data System (ADS) under Cooperative Agreement 80NSSC21M00561.
    We further acknowledge the Python programming language and its scientific ecosystem for enabling data analysis and visualization.
    This work used the following libraries: Astropy~\citep{Astropy+2013, Astropy+2018, Astropy+2022}, NumPy~\citep{Harris+2020}, Matplotlib~\citep{Hunter2007}, and SciPy~\citep{Virtanen+2020}, as well as the \texttt{Euclid\_tools} \citep{Kiwy+2025} Python package for retrieving, inspecting, and classifying Euclid spectral data.
    During the preparation of this work, the author used ChatGPT to improve the readability and language of the manuscript.

    \bibliography{manuscript}{}
    \bibliographystyle{aasjournalv7}

    \appendix

    \section{Spectral Characteristics of Candidates}\label{sec:spectral-characteristics-of-candidates}

    In the following, we describe the spectral characteristics of each candidate, focusing on their alignment with the \textbf{Burgasser templates} (Figure~\ref{fig:comparison-with-burgasser-templates}) and the reliability of their classifications.
    The spectral types listed in parentheses after each object name correspond to those inferred from the Burgasser templates; the final adopted spectral types, obtained as the mean of the Burgasser and Theissen classifications, are summarized in Table~\ref{tab:photometric-distance-table}.\\

    \textbf{E0322\ms2650\,(T7)}
    Elevated noise is present across the full wavelength range of the spectrum.
    While the $J$-band peak aligns reasonably well with the T7 standard, the $H$-band peak appears slightly elevated relative to the template.
    The flux within the $J$-band methane and water absorption troughs is somewhat lower than that of the T7 standard.
    Despite these discrepancies, the overall spectral morphology remains consistent with that of a late-T dwarf.

    \textbf{E0328\ms2749\,(T4)}
    Originally reported as a T6 by \citet{Dominguez-Tagle+2025} and noted as pecular in~\cite{Zerjal+2025}, this object shows a moderately noisy spectrum with noticeable scatter in the $J$-band peak.
    Similar to E0322\ms2650, it exhibits an elevated $H$-band peak; however, the overall spectral shape remains well defined.
    Although the flux values between 1.3 and \SI{1.6}{\micro\metre} are flagged as problematic (orange line), the spectrum still captures the essential features of a mid-T dwarf.

    \textbf{E0329\ms2901\,(T2)}
    Previously classified as a T4 by \citet{Dominguez-Tagle+2025}, this spectrum shows clear signatures of an early-T dwarf despite moderate noise.
    Comparable to E0328\ms2749, the spectrum shows elevated flux in the $H$-band relative to the standard, along with increased scatter in the $J$-band peak.
    Nonetheless, the overall spectral shape remains broadly consistent with our T2 classification.

    \textbf{E0331\ms2618\,(T4)}
    Despite the elevated noise, the spectrum of this object—classified as T5 by \citet{Dominguez-Tagle+2025}—shows good overall agreement with the T4 template.
    The $J$-band peak is well defined, and the $H$-band displays a clear methane absorption feature.
    Overall, the spectral morphology is consistent with a mid-T classification, though the noise level introduces some uncertainty in the subtype assignment.

    \textbf{E0331\ms2725\,(T4)}
    This spectrum is notably noisy, which limits the precision of the spectral classification.
    While the $J$-band peak is reasonably well aligned with the T4 standard, significant deviations are observed in the methane and water absorption features.
    The $J$-band water absorption trough shows excess flux relative to the template, whereas the $H$-band methane trough dips well below the expected level.
    Additionally, the $H$-band peak is also not well defined.
    These discrepancies warrant caution in interpreting the spectral type.

    \textbf{E0331\ms2712\,(T4)}
    Substantial uncertainty across the entire wavelength range affects the reliability of this spectrum.
    The $J$-band peak near \SI{1.3}{\micro\metre} appears elevated relative to the T4 standard.
    The $H$-band peak and methane absorption trough also show deviations from the template, likely due to the low signal-to-noise ratio.
    Despite these limitations, the overall spectral morphology is broadly consistent with a mid-T dwarf classification, although the precise subtype should be interpreted with caution.

    \textbf{E0348\ms5016\,(T3)}
    This object has been reported by \citet{Zerjal+2025} as a T dwarf without a specific subtype.
    Its spectrum is among the cleaner examples in the sample, although the $J$-band peak shows some scatter.
    Between 1.33 and \SI{1.5}{\micro\metre}, the observed flux slightly exceeds that of the T3 template.
    Beyond \SI{1.5}{\micro\metre}, the spectrum closely follows the expected shape for a T3 dwarf, with a decently aligned $H$-band peak and trough.
    Overall, the fit is acceptable, supporting the assigned spectral type.

    \textbf{E0352\ms4910\,(T6)}
    This object was previously identified as a T7 by both \citet{JYZhang+2024} and \citet{Dominguez-Tagle+2025}.
    Its spectrum is clean and smooth throughout, with low noise and a stable continuum across the full wavelength range.
    It closely matches the T6 template, with particularly well-aligned methane and water absorption features.
    A slightly elevated $H$-band peak suggests a marginally later subtype (T7–T8), but the overall morphology supports a confident mid-to-late T classification.

    \textbf{E0359\ms4740\,(T4)}
    Reported as a T8 dwarf by \citet{JYZhang+2024} and as T7 by \citet{Dominguez-Tagle+2025}, this source exhibits a relatively smooth spectrum despite modest scatter in the $H$-band absorption troughs.
    Both the $J$- and $H$-band peaks appear elevated relative to the T4 standard, though the water absorption troughs between these peaks align well with the template.
    Given the discrepancies in peak amplitudes, the T4 classification should be considered with caution.

    \textbf{E0403\ms4725\,(T5)}
    The spectrum exhibits moderate noise throughout, which impacts the clarity of key features.
    Notable deviations from the T5 template are observed, particularly in the $J$-band absorption troughs, where the flux differs from the expected morphology.
    Although the overall structure is broadly consistent with a mid-T dwarf, the limited spectral quality constrains the precision of the type assignment.

    \textbf{E0404\ms4639\,(T4)}
    This is one of the noisiest spectra in the sample, with a significantly degraded signal in both the $J$- and $H$-band regions.
    The $J$-band peak aligns reasonably well with the T4 standard, but the absorption features between 1.3 and \SI{1.5}{\micro\metre} fall slightly below the expected flux level.
    Additionally, the $H$-band peak and trough are poorly defined, likely due to the low signal-to-noise ratio.
    While the overall shape is not inconsistent with a mid-T dwarf, the high noise level reduces confidence in the T4 classification, which should be considered tentative.

    \textbf{E0418\ms4804\,(T6)}
    Despite a relatively noisy continuum, the overall spectral morphology is broadly consistent with the T6 standard.
    The $J$-band peak is clearly defined, though slightly broader than in the template.
    A noticeable flux deficit is present between 1.6 and \SI{1.65}{\micro\metre}, affecting the match in the $H$-band methane absorption feature.
    Nonetheless, the key features are recognizable, and the classification as a mid-to-late T dwarf is supported, albeit with moderate confidence.

    \textbf{E1745\ps6459\,(T4)}
    This object was previously classified as T7 by \citet{Mace+2013}, \citet{JYZhang+2024} and as T6 by \citet{Dominguez-Tagle+2025}.
    It is among the higher-quality spectra in our sample, showing a strong match to the T4 template.
    Both the $J$- and $H$-band peaks are well aligned in amplitude and shape, and the overall continuum tracks the standard closely.
    Minor deviations in the $H$-band methane absorption do not affect the mid-T classification.

    \textbf{E1747\ps6454\,(T2)}
    Despite a noise level only slightly higher than that of E1745\ps6459, this spectrum shows notable discrepancies with the T2 template.
    Key spectral regions in the $J$- and $H$-bands are difficult to interpret due to deviations in both peak amplitude and absorption troughs.
    While the overall flux trend follows the expected pattern for an early-T dwarf, the mismatch in specific features reduces the confidence in the assigned classification.

    \textbf{E1805\ps6444\,(T2)}
    This spectrum exhibits significant noise, particularly in the $J$-band, which affects the clarity of key absorption features.
    The flux in the overlapping region of methane and water absorption within the $J$-band is noticeably lower than expected for a T2 dwarf, while the $H$-band peak appears elevated relative to the standard.
    Although the general spectral morphology aligns with an early-T classification, the combination of noise and discrepancies in diagnostic regions limits confidence in the precise subtype.

    \begin{figure*}
        \section{Comparison with Theissen Templates}\label{sec:comparison-with-theissen-templates}
        \begin{center}
            \includegraphics[width=15cm]{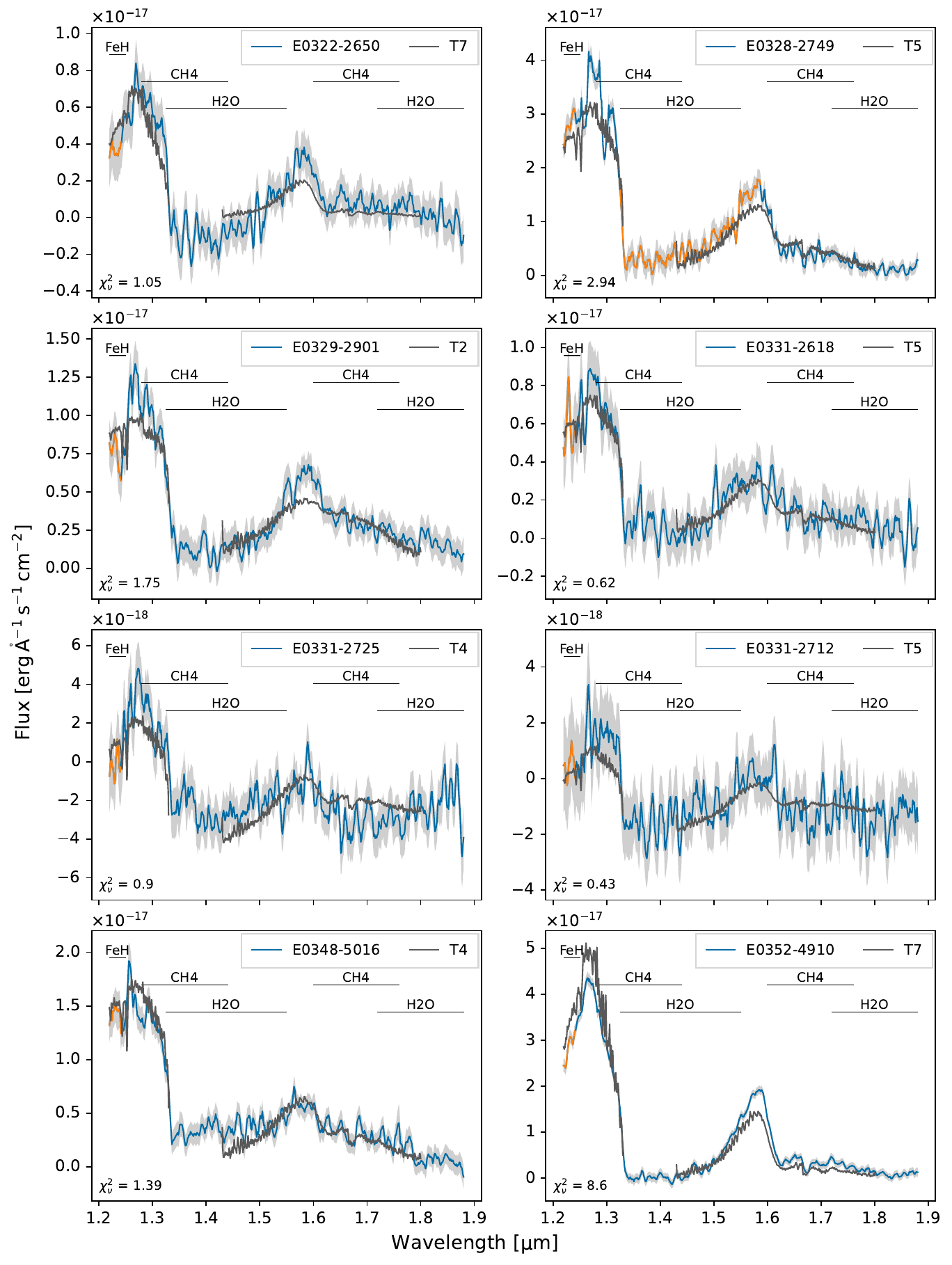}
        \end{center}
        \caption{
            Comparison of smoothed candidate spectra (blue curves) with the \textbf{Theissen templates} (dark gray curves) for the 15 selected candidates. As shown in Figure~\ref{fig:comparison-with-burgasser-templates}, the orange curves denote flagged values, while the gray shading indicates the flux uncertainty. Key molecular absorption features are marked by horizontal black lines. The inferred spectral types based on the Theissen templates are generally consistent with those derived from the Burgasser templates and agree within one subtype.
        }
        \label{fig:comparison-with-theissen-templates}
    \end{figure*}

    \begin{figure*}
        \begin{center}
            \includegraphics[width=15cm]{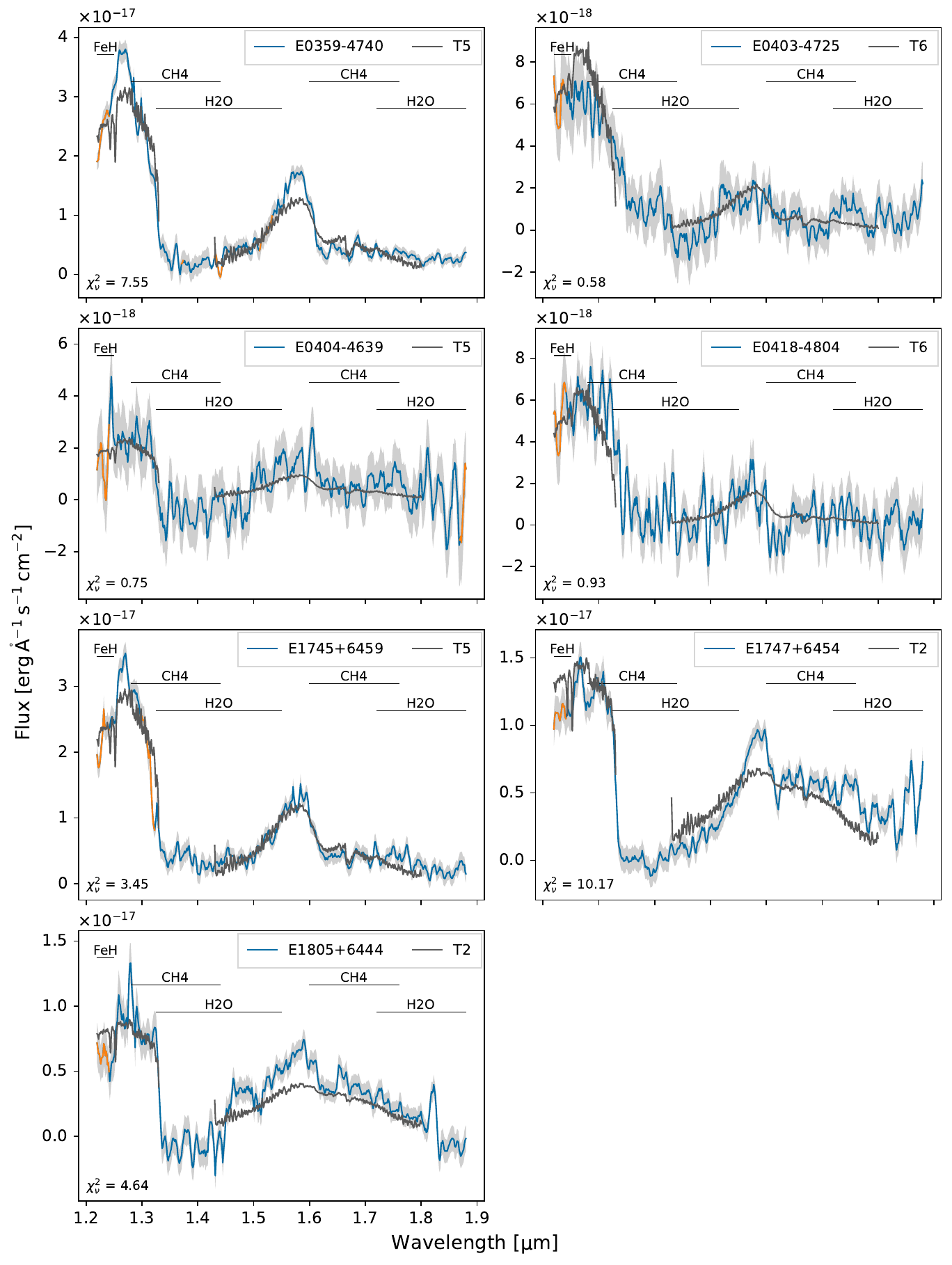}
        \end{center}
        \centering\textbf{Figure~\ref{fig:comparison-with-theissen-templates}.} (Continued)
    \end{figure*}

    \clearpage
    \startlongtable
    \begin{deluxetable*}{lrlrrrrrr}
        \section{Synthetic Absolute Magnitudes and Colors}\label{sec:synthetic-absolute-magnitudes-and-colors}
        \tablecaption{Synthetic absolute magnitudes and colors of ultracool dwarfs in the Euclid NISP bands \label{tab:abs-mag-table}}
        \tablehead{
            \colhead{SpT} &
            \colhead{Num. SpT} &
            \colhead{Age} &
            \colhead{$M_Y$} &
            \colhead{$M_J$} &
            \colhead{$M_H$} &
            \colhead{$Y - J$} &
            \colhead{$Y - H$} &
            \colhead{$J - H$} \\
            \colhead{~} &
            \colhead{~} &
            \colhead{~} &
            \colhead{(mag)} &
            \colhead{(mag)} &
            \colhead{(mag)} &
            \colhead{(mag)} &
            \colhead{(mag)} &
            \colhead{(mag)}
        }
        \startdata
        M6 & 6  & field & 11.634 & 11.321 & 11.288 & 0.313 & 0.346  & 0.033  \\
        M6 & 6  & young & 10.067 & 9.804  & 9.800  & 0.263 & 0.267  & 0.004  \\
        M7 & 7  & field & 11.869 & 11.574 & 11.480 & 0.295 & 0.389  & 0.094  \\
        M7 & 7  & young & 10.912 & 10.636 & 10.542 & 0.276 & 0.370  & 0.094  \\
        M8 & 8  & field & 12.167 & 11.851 & 11.718 & 0.316 & 0.449  & 0.133  \\
        M8 & 8  & young & 11.617 & 11.298 & 11.141 & 0.319 & 0.476  & 0.157  \\
        M9 & 9  & field & 12.522 & 12.162 & 11.996 & 0.360 & 0.526  & 0.166  \\
        M9 & 9  & young & 12.245 & 11.859 & 11.650 & 0.386 & 0.595  & 0.209  \\
        L0 & 10 & field & 12.925 & 12.509 & 12.305 & 0.416 & 0.620  & 0.204  \\
        L0 & 10 & young & 12.835 & 12.371 & 12.107 & 0.464 & 0.728  & 0.264  \\
        L1 & 11 & field & 13.361 & 12.887 & 12.636 & 0.474 & 0.725  & 0.251  \\
        L1 & 11 & young & 13.410 & 12.862 & 12.535 & 0.548 & 0.875  & 0.327  \\
        L2 & 12 & field & 13.813 & 13.285 & 12.978 & 0.528 & 0.835  & 0.307  \\
        L2 & 12 & young & 13.977 & 13.348 & 12.949 & 0.629 & 1.028  & 0.399  \\
        L3 & 13 & field & 14.262 & 13.691 & 13.320 & 0.571 & 0.942  & 0.371  \\
        L3 & 13 & young & 14.531 & 13.828 & 13.352 & 0.703 & 1.179  & 0.476  \\
        L4 & 14 & field & 14.689 & 14.088 & 13.650 & 0.601 & 1.039  & 0.438  \\
        L4 & 14 & young & 15.060 & 14.294 & 13.741 & 0.766 & 1.319  & 0.553  \\
        L5 & 15 & field & 15.077 & 14.462 & 13.961 & 0.615 & 1.116  & 0.501  \\
        L5 & 15 & young & 15.543 & 14.731 & 14.109 & 0.812 & 1.434  & 0.622  \\
        L6 & 16 & field & 15.410 & 14.796 & 14.244 & 0.614 & 1.166  & 0.552  \\
        L6 & 16 & young & 15.959 & 15.120 & 14.446 & 0.839 & 1.513  & 0.674  \\
        L7 & 17 & field & 15.676 & 15.078 & 14.495 & 0.598 & 1.181  & 0.583  \\
        L7 & 17 & young & 16.290 & 15.440 & 14.742 & 0.850 & 1.548  & 0.698  \\
        L8 & 18 & field & 15.869 & 15.299 & 14.712 & 0.570 & 1.157  & 0.587  \\
        L8 & 18 & young & 16.517 & 15.674 & 14.989 & 0.843 & 1.528  & 0.685  \\
        L9 & 19 & field & 15.988 & 15.458 & 14.899 & 0.530 & 1.089  & 0.559  \\
        L9 & 19 & young & 16.632 & 15.812 & 15.181 & 0.820 & 1.451  & 0.631  \\
        T0 & 20 & field & 16.042 & 15.557 & 15.062 & 0.485 & 0.980  & 0.495  \\
        T0 & 20 & young & 16.636 & 15.851 & 15.322 & 0.785 & 1.314  & 0.529  \\
        T1 & 21 & field & 16.046 & 15.610 & 15.215 & 0.436 & 0.831  & 0.395  \\
        T1 & 21 & young & 16.543 & 15.800 & 15.420 & 0.743 & 1.123  & 0.380  \\
        T2 & 22 & field & 16.027 & 15.638 & 15.378 & 0.389 & 0.649  & 0.260  \\
        T2 & 22 & young & 16.385 & 15.683 & 15.497 & 0.702 & 0.888  & 0.186  \\
        T3 & 23 & field & 16.023 & 15.678 & 15.576 & 0.345 & 0.447  & 0.102  \\
        T3 & 23 & young & 16.213 & 15.543 & 15.586 & 0.670 & 0.627  & \ms0.043 \\
        T4 & 24 & field & 16.087 & 15.775 & 15.843 & 0.312 & 0.244  & \ms0.068 \\
        T4 & 24 & young & 16.099 & 15.444 & 15.734 & 0.655 & 0.365  & \ms0.290 \\
        T5 & 25 & field & 16.282 & 15.993 & 16.221 & 0.289 & 0.061  & \ms0.228 \\
        T5 & 25 & young & 16.144 & 15.473 & 16.007 & 0.671 & 0.137  & \ms0.534 \\
        T6 & 26 & field & 16.691 & 16.409 & 16.763 & 0.282 & \ms0.072 & \ms0.354 \\
        T6 & 26 & young & 16.478 & 15.747 & 16.488 & 0.731 & \ms0.010 & \ms0.741 \\
        T7 & 27 & field & 17.411 & 17.120 & 17.529 & 0.291 & \ms0.118 & \ms0.409 \\
        T7 & 27 & young & 17.261 & 16.411 & 17.284 & 0.850 & \ms0.023 & \ms0.873 \\
        T8 & 28 & field & 18.559 & 18.240 & 18.590 & 0.319 & \ms0.031 & \ms0.350 \\
        T8 & 28 & young & 18.692 & 17.646 & 18.523 & 1.046 & 0.169  & \ms0.877 \\
        T9 & 29 & field & 20.270 & 19.906 & 20.030 & 0.364 & 0.240  & \ms0.124 \\
        T9 & 29 & young & 21.004 & 19.667 & 20.360 & 1.337 & 0.644  & \ms0.693 \\
        \enddata
        \tablecomments{
            Synthetic absolute magnitudes ($M_Y$, $M_J$, $M_H$) and colors ($Y - J$, $Y - H$, $J - H$) in the AB magnitude system for spectral types M6–T9 in the Euclid photometric system. These values are derived from the polynomial relations of \citet{Sanghi+2024} and are provided separately for field-age and young-age populations. Spectral types are listed in both standard notation (SpT) and as numerical codes (Num. SpT). The synthetic sequences are used in our analysis for photometric type and distance estimations of the candidate brown dwarfs.
        }
    \end{deluxetable*}

    \begin{deluxetable*}{llccccccccl}
        \section{Detections in Other Surveys}\label{sec:detections-in-other-surveys}
        \tablecaption{Near- and mid-infrared photometry from VHS, VIKING, and CatWISE2020 for detected candidates \label{tab:vhs_viking_wise_phot}}
        \tablehead{
            \colhead{~} &
            \multicolumn{6}{c|}{Near-infrared photometry} &
            \multicolumn{4}{c}{Mid-infrared photometry} \\
            \cline{2-7} \cline{8-11}
            \colhead{Object} &
            \colhead{Survey} &
            \colhead{$z$} &
            \colhead{$Y$} &
            \colhead{$J$} &
            \colhead{$H$} &
            \colhead{$K_s$} &
            \colhead{$W1$} &
            \colhead{$W2$} &
            \colhead{$W1 - W2$} &
            \colhead{Type} \\
            \colhead{~} &
            \colhead{~} &
            \colhead{(mag)} &
            \colhead{(mag)} &
            \colhead{(mag)} &
            \colhead{(mag)} &
            \colhead{(mag)} &
            \colhead{(mag)} &
            \colhead{(mag)} &
            \colhead{(mag)} &
            \colhead{~}
        }
        \startdata
        E0328\ms2749 & VIKING & 22.041$\pm$0.169 & 19.255$\pm$0.034 & 17.839$\pm$0.015 & 18.167$\pm$0.053 & 17.872$\pm$0.066 & 17.527$\pm$0.078 & 16.228$\pm$0.068 & 1.299 & T4 \\
        E0329\ms2901 & VIKING & \ldots & 20.714$\pm$0.101 & 19.311$\pm$0.051 & 19.446$\pm$0.161 & 18.966$\pm$0.167 & \ldots & \ldots & \ldots & \ldots \\
        E0331\ms2618 & VHS    & \ldots & \ldots & 19.798$\pm$0.168 & \ldots & \ldots & \ldots & \ldots & \ldots & \ldots \\
        E0331\ms2725 & VIKING & \ldots & \ldots & 20.758$\pm$0.194 & \ldots & \ldots & \ldots & \ldots & \ldots & \ldots \\
        E0331\ms2712 & VIKING & \ldots & 21.712$\pm$0.278 & 20.615$\pm$0.171 & \ldots & \ldots & \ldots & \ldots & \ldots & \ldots \\
        E0348\ms5016 & VHS    & \ldots & \ldots & 18.962$\pm$0.078 & 18.893$\pm$0.162 & \ldots & \ldots & \ldots & \ldots & \ldots \\
        E0352\ms4910 & VHS    & \ldots & \ldots & 17.832$\pm$0.036 & 18.218$\pm$0.123 & 18.170$\pm$0.169 & 17.395$\pm$0.051 & 15.264$\pm$0.024 & 2.131 & T6 \\
        E0359\ms4740 & VHS    & \ldots & \ldots & 18.073$\pm$0.031 & 18.291$\pm$0.062 & 18.508$\pm$0.188 & 17.231$\pm$0.051 & 15.267$\pm$0.027 & 1.964 & T5 \\
        E0403\ms4725 & VHS    & \ldots & \ldots & 19.985$\pm$0.146 & \ldots & \ldots & \ldots & \ldots & \ldots & \ldots \\
        E0404\ms4639 & VHS    & \ldots & \ldots & 20.266$\pm$0.183 & \ldots & \ldots & \ldots & \ldots & \ldots & \ldots \\
        E0418\ms4804 & VHS    & \ldots & \ldots & 20.215$\pm$0.208 & \ldots & \ldots & \ldots & \ldots & \ldots & \ldots \\
        E1745\ps6459 & \ldots & \ldots & \ldots & \ldots & \ldots & \ldots & 18.032$\pm$0.050 & 16.240$\pm$0.033 & 1.792 & T5 \\
        E1747\ps6454 & \ldots & \ldots & \ldots & \ldots & \ldots & \ldots & 18.325$\pm$0.058 & 16.861$\pm$0.040 & 1.464 & T4 \\
        E1805\ps6444 & \ldots & \ldots & \ldots & \ldots & \ldots & \ldots & 19.004$\pm$0.083 & 17.453$\pm$0.056 & 1.551 & T4 \\
        \enddata
        \tablecomments{
            VHS (DR7) and VIKING (DR5) $zYJHK_s$ photometry is reported where available, supplemented by CatWISE2020 mid-infrared $W1$ and $W2$ measurements.
            $W1-W2$ colors are used to infer photometric types based on the empirical relations of \citet{Best+2018}, generally consistent with our adopted spectral classifications listed in Table~\ref{tab:photometric-distance-table}. All magnitudes are on the Vega system. Missing values are indicated by an ellipsis.
        }
    \end{deluxetable*}

\end{document}